\documentclass[prb,twocolumn,notitlepage,superscriptaddress,showkeys,showpacs]{revtex4-1}
\usepackage{amsmath,amsfonts,amssymb,amsthm,epsfig,array}
\usepackage{dsfont}
\usepackage{graphics}
\usepackage{titletoc}
\usepackage{titlesec}
\usepackage{float}
\usepackage{verbatim}
\usepackage[dvipsnames]{xcolor}
\usepackage[hidelinks,colorlinks,linkcolor=blue,
citecolor=blue,urlcolor=blue]{hyperref}
\usepackage[titletoc,title]{appendix}
\usepackage{slashed}

\newcommand{\bsl}[1]{\boldsymbol{#1}}
\newcommand{\shpa}{\shortparallel}

\newcommand{\eqnref}[1]{Eq.\,\eqref{#1}}
\newcommand{\figref}[1]{Fig.\,\ref{#1}}
\newcommand{\tabref}[1]{Tab.\,\ref{#1}}
\newcommand{\secref}[1]{Sec.\,\ref{#1}}

\newcommand{\refcite}[1]{Ref.\,[\onlinecite{#1}]}

\newcommand{\eq}[1]{\begin{equation} #1 \end{equation}}
\newcommand{\eqa}[1]{\begin{align}\begin{split} #1 \end{split}\end{align}}
\let\oldAA\AA
\renewcommand{\AA}{\text{\normalfont\oldAA}}

\begin{document}
\title{Surface Majorana Flat Bands in $j=\frac{3}{2}$ Superconductors with Singlet-Quintet Mixing}
\author{Jiabin Yu}
\affiliation{Department of Physics, the Pennsylvania State University, University Park, PA, 16802}
\author{Chao-Xing Liu}
\email{cxl56@psu.edu}
\affiliation{Department of Physics, the Pennsylvania State University, University Park, PA, 16802}
\begin{abstract}
Recent experiments\cite{kim2016YPtBiSCj=3/2} have revealed the evidence of nodal-line superconductivity in half-Heusler superconductors, e.g. YPtBi. Theories have suggested the topological nature of such nodal-line superconductivity and proposed the existence of surface Majorana flat bands on the (111) surface of half-Heusler superconductors.
Due to the divergent density of states of the surface Majorana flat bands, the surface order parameter and the surface impurity play essential roles in determining the surface properties.
In this work, we studied the effect of the surface order parameter and the surface impurity on the surface Majorana flat bands of half-Heusler superconductors based on the Luttinger model.
To be specific, we consider the topological nodal-line superconducting phase induced by the singlet-quintet pairing mixing, classify all the possible translationally invariant order parameters for the surface states according to irreducible representations of $C_{3v}$ point group, and demonstrate that any energetically favorable order parameter needs to break time-reversal symmetry.
We further discuss the energy splitting in the energy spectrum of surface Majorana flat bands induced by different order parameters and non-magnetic or magnetic impurities.
We proposed that the splitting in the energy spectrum can serve as the fingerprint of the pairing symmetry and mean-field order parameters.
Our theoretical prediction can be examined in the future scanning tunneling microscopy experiments.
\end{abstract}
\maketitle

\section{Introduction}
Recent years have witnessed increasing research interests in half-Heusler compounds ($R$PdBi or $R$PtBi with $R$ a rare-earth element)\cite{Graf2011heusler} due to their non-trivial band topology\cite{Lin2010Half,Chadov2010tunable,Xiao2010Half,
AlSawai2010Half,Yan2014half,Liu2016halfSS,Logan2016SShalf,
Cano2016chiral,Ruan2016WSM,Hirschberger2016WSM,
Shekhar2016observationCMT,Suzuki2016largeAHE,Yang2017HHTP,
Liu2018TI}, magnetism\cite{Pan2013ErPdBiSC,Gofryk2011Mag,Muller2014Mag,
Nikitin2015HoPdBiSC,Nakajima2015RPdBiSC,Pavlosiuk2016AFMSCHH, Pavlosiuk2016MagHH,Yu2017ModelAFMHH,Pavlosiuk2018MagHH}
and unconventional superconductivity\cite{Goll2008LaBiPtSC,Butch2011SCYPtBi,
Bay2012SCYPtBi,kim2016YPtBiSCj=3/2,Tafti2013LuPtBiSC,
Pan2013ErPdBiSC,Nakajima2015RPdBiSC,Xu2014LuPdBiSC,
Pavlosiuk2015LuPdBiSC,Nikitin2015HoPdBiSC,
Meinert2015UnconverntialSCYPtBi,Pavlosiuk2016AFMSCHH,
TbPdBi2018SC}.
Half-Heusler superconductors (SCs) are of particular interest because of the low carrier density ($10^{18}\sim 10^{19} cm^{-3}$), the power-law temperature dependence of London penetration depth, and the large upper critical field.
Furthermore, it was theoretically proposed that electrons near Fermi level in half-Heusler SCs possess total angular momentum $j=\frac{3}{2}$ as a result of the addition of the $\frac{1}{2}$ spin and the angular momentum of p atomic orbitals ($l=1$).\cite{kim2016YPtBiSCj=3/2,Brydon2016j=3/2SC}
Therefore, half-Heusler SCs provide a great platform to study the superconductivity of $j=\frac{3}{2}$ fermions.
Such $j=\frac{3}{2}$ fermions were also studied in anti-perovskite materials\cite{Kawakami2018j=3/2electrons} and the cold atom system\cite{Wu2006spin3/2CAS,Kuzmenko2018F=3/2CFG}.
Due to the $j=\frac{3}{2}$ nature, the spin of Cooper pairs can take four values: $S=0$ (singlet), 1 (triplet), 2 (quintet) and 3 (septet), among which quintet and septet Cooper pairs cannot appear for spin-$\frac{1}{2}$ electrons.

In order to understand the unconventional superconductivity, various pairing states were proposed, including mixed singlet-septet pairing\cite{Brydon2016j=3/2SC,kim2016YPtBiSCj=3/2,
Yang2017Majoranaj=3/2SC,Timm2017nodalj=3/2SC},
mixed singlet-quintet pairing\cite{yu2017Singlet-Quintetj=3/2SC,
Wang2018j=3/2SCSurface,Yu2018SSUCFDE}, s-wave quintet pairing
\cite{Brydon2016j=3/2SC,Roy2017j=3/2SC,Timm2017nodalj=3/2SC,
Boettcher2018j=3/2SC}
, d-wave quintet pairing\cite{Yang2016j=3/2Fermions,Venderbos2018j=3/2SC}
, odd-parity (triplet and septet) parings\cite{Yang2016j=3/2Fermions,Venderbos2018j=3/2SC,
Savary2017j=3/2SC,Ghorashi2017j=3/2SCdisorder}, {\it et al}\cite{Venderbos2018j=3/2SC,Brydon2018BFS}.
In particular, \refcite{Brydon2016j=3/2SC,kim2016YPtBiSCj=3/2,
Yang2017Majoranaj=3/2SC,Timm2017nodalj=3/2SC,yu2017Singlet-Quintetj=3/2SC,
Wang2018j=3/2SCSurface,Yu2018SSUCFDE} proposed that the power-law temperature dependence of London penetration depth can be explained by topological nodal-line superconductivity (TNLS) generated by the pairing mixing between different spin channels.
In particular, it has been shown that two types of pairing mixing states, the singlet-quintet mixing and singlet-septet mixing, can both give rise to nodal lines in certain parameter regimes.

In this work, we focus on the singlet-quintet mixing, which was proposed in \refcite{yu2017Singlet-Quintetj=3/2SC}.
As a consequence of TNLS, the Majorana flat bands (MFBs) are expected to exist on the surface perpendicular to certain directions.
Such surface MFBs (SMFBs) are expected to show divergent quasi-particle density of states (DOS) at the Fermi energy and thus can be directly probed through experimental techniques, such as scanning tunneling microscopy (STM). \cite{Yada2011SDOSTSC}
Due to the divergent DOS, certain types of interaction \cite{Li2013MZMJC,Potter2014EdgeMZM,
Timm2015SurfInsNodalSC,Hofmann2016EdgeMZMIns} and surface impurities\cite{Ikegaya2015APE, Ikegaya2017MZMDirtyNSC,Ikegaya2018SymABSDirty} are expected to have a strong influence on SMFBs.
This motivates us to study the effect of the interaction-induced surface order parameter and the surface impurity on the SMFBs of the superconducting Luttinger model with the singlet-quintet mixing.
Specifically, we classify all the mean-field translationally invariant order parameters of the SMFBs according to the irreducible representations (IRs) of $C_{3v}$ group, identify their possible physical origins, and show their energy spectrum by calculating the corresponding DOS.
We find that the order parameter needs to break the time-reversal (TR) symmetry in order to either gap out the SMFBs or convert the SMFBs to nodal-lines or nodal points.
We also study the quasi-particle local DOS (LDOS) of SMFBs with a surface charge impurity or a surface magnetic impurity (whose magnetic moment is perpendicular to the surface), and show that the peak splitting induced by different types of impurities can help to distinguish the pairing symmetries and surface order parameters.

The rest of the paper is organized as the following.
In \secref{sec:model_H} and \ref{sec:surf_MFB}, we briefly review the superconducting Luttinger model with singlet-quintet mixing and illustrate the symmetry properties of SMFBs.
In \secref{sec:MF_order_MFB}, we classify all the mean-field translationally invariant order parameters according to the IRs of $C_{3v}$ and identify their physical origin.
We also calculate the energy spectrum and DOS of SMFBs with different order parameters.
In \secref{sec:imp_MFB}, the impurity effect on the LDOS of MFBs with/without the surface order parameter is discussed.
Finally, our work is concluded in \secref{sec:conclusion}

\section{Model Hamiltonian}
\label{sec:model_H}
The model that we used to generate MFBs in this work is the same as that studied in \refcite{yu2017Singlet-Quintetj=3/2SC}, which describes the superconductivity in the Luttinger model with mixed s-wave singlet and isotropic d-wave quintet channels.
The Bogoliubov-de-Gennes (BdG) Hamiltonian in the continuous limit reads
\begin{equation}
\label{eq:H_BdG}
H=\frac{1}{2}\sum_{\bsl{k}}\Psi_{\bsl{k}}^{\dagger}h_{BdG}(\bsl{k})\Psi_{\bsl{k}}+const.\ ,
\end{equation}
where $\Psi^{\dagger}_{\bsl{k}}=(c_{\bsl{k}}^{\dagger},c_{-\bsl{k}}^{T})$ is the Nambu spinor and
$c_{\bsl{k}}^{\dagger}=(c_{\bsl{k},\frac{3}{2}}^{\dagger},c_{\bsl{k},\frac{1}{2}}^{\dagger},c_{\bsl{k},-\frac{1}{2}}^{\dagger},c_{\bsl{k},-\frac{3}{2}}^{\dagger})$
are creation operators of $j=\frac{3}{2}$ fermionic excitations.
The term
\begin{equation}
h_{BdG}(\bsl{k})=
\left(
\begin{matrix}
h(\bsl{k})& \Delta(\bsl{k})\\
\Delta^{\dagger}(\bsl{k})& -h^T(-\bsl{k})\\
\end{matrix}
\right)
\end{equation}
consists of the normal part $h(\bsl{k})$ that is the Luttinger model\cite{Luttinger1956LuttingerModel,Chadov2010tunable,
Winkler2003SOC, yu2017Singlet-Quintetj=3/2SC}
\begin{eqnarray}\label{Eqn:h}
h(\bsl{k})
=(\frac{k^2}{2m}-\mu)\Gamma^0+
c_1 \sum_{i=1}^{3}g_{\bsl{k},i}\Gamma^i+c_2 \sum_{i=4}^{5}g_{\bsl{k},i}\Gamma^i
\end{eqnarray}
and the paring part $\Delta(\bsl{k})$ that contains s-wave singlet and isotropic d-wave quintet channels
\begin{equation}
\label{eq:pairing}
\Delta(\bsl{k})
=\Delta_0\frac{\Gamma^0}{2}\gamma
+
\Delta_1\sum_{i=1}^5 \frac{a^2 g_{\bsl{k},i}\Gamma^i}{2}\gamma,
\end{equation}
where $\mu$ is the chemical potential,
$c_1,c_2$ indicate the strength of the centrosymmetric spin orbital coupling (SOC) which is the coupling between the orbital and the 3/2-``spin'',
d-wave cubic harmonics $g_{\bsl{k},i}$ and five $\Gamma$ matrices are shown in Appendix.\ref{app:conv_expn},
$\Delta_{0,1}$ are order parameters of singlet and quintet channels, respectively,
$a$ is the lattice constant of the material, and
$\gamma=-\Gamma^1\Gamma^3$ is the TR matrix.
The coexistence of the two order parameters is allowed by their same symmetry properties~\cite{yu2017Singlet-Quintetj=3/2SC,Blount1985SC,Ueda1985SC,Volovik1985SC,
Sigrist1991SC,Annett1990SC,Annett1991SC,Annett1996SC}.

Before demonstrating the SMFB generated by \eqnref{eq:H_BdG}, we first discuss the symmetry properties of the Hamiltonian $H$.
As discussed in \refcite{yu2017Singlet-Quintetj=3/2SC}, $H$ has TR symmetry, and its point group is $O(3)$ or $O_h$ for $c_1=c_2$ or $c_1\neq c_2$, respectively.
Due to the coexistence of TR and inversion symmetries, the Luttinger model $h(\bsl{k})$ has two doubly degenerate bands $\xi_{\pm}(\bsl{k})=k^2/(2m_{\pm})-\mu$,
where $m_{\pm}=m \widetilde{m}_{\pm}$ are effective masses of two bands, $\widetilde{m}_{\pm}=1/(1\pm 2mQ_c)$,  $Q_c=\sqrt{c_1^2 Q_1^2+c_2^2 Q_2^2}$, $Q_1=\sqrt{\hat{g}^2_{1}+\hat{g}^2_{2}+\hat{g}^2_{3}}$, $Q_2=\sqrt{\hat{g}^2_{4}+\hat{g}^2_{5}}$, and $\hat{g}_i=g_i/k^2$.
In addition, particle-hole (PH) symmetry can be defined as $-\mathcal{C} h^*_{BdG}(-\bsl{k})\mathcal{C}^{\dagger}=h_{BdG}(\bsl{k}) $ and $\Psi_{\bsl{k}}^{\dagger} \mathcal{C}=\Psi_{-\bsl{k}}^T$ for the BdG Hamiltonian, where
$\mathcal{C}=\tau_x$ with $\tau_x$ the Pauli matrix for the PH index.
Combining the PH and TR symmetries, we have the chiral symmetry $-\chi h_{BdG}(\bsl{k}) \chi^{\dagger}=h_{BdG}(\bsl{k})$,
where $\chi=i \mathcal{T C^*}$ and $\mathcal{T}=\text{diag}(\gamma,\gamma^*)$ is the TR matrix on the Nambu bases.
The representations of other symmetry operators are shown in Appendix.\ref{app:rep_sym}.

\begin{figure}[t]
    \centering
    \includegraphics[width=\columnwidth]{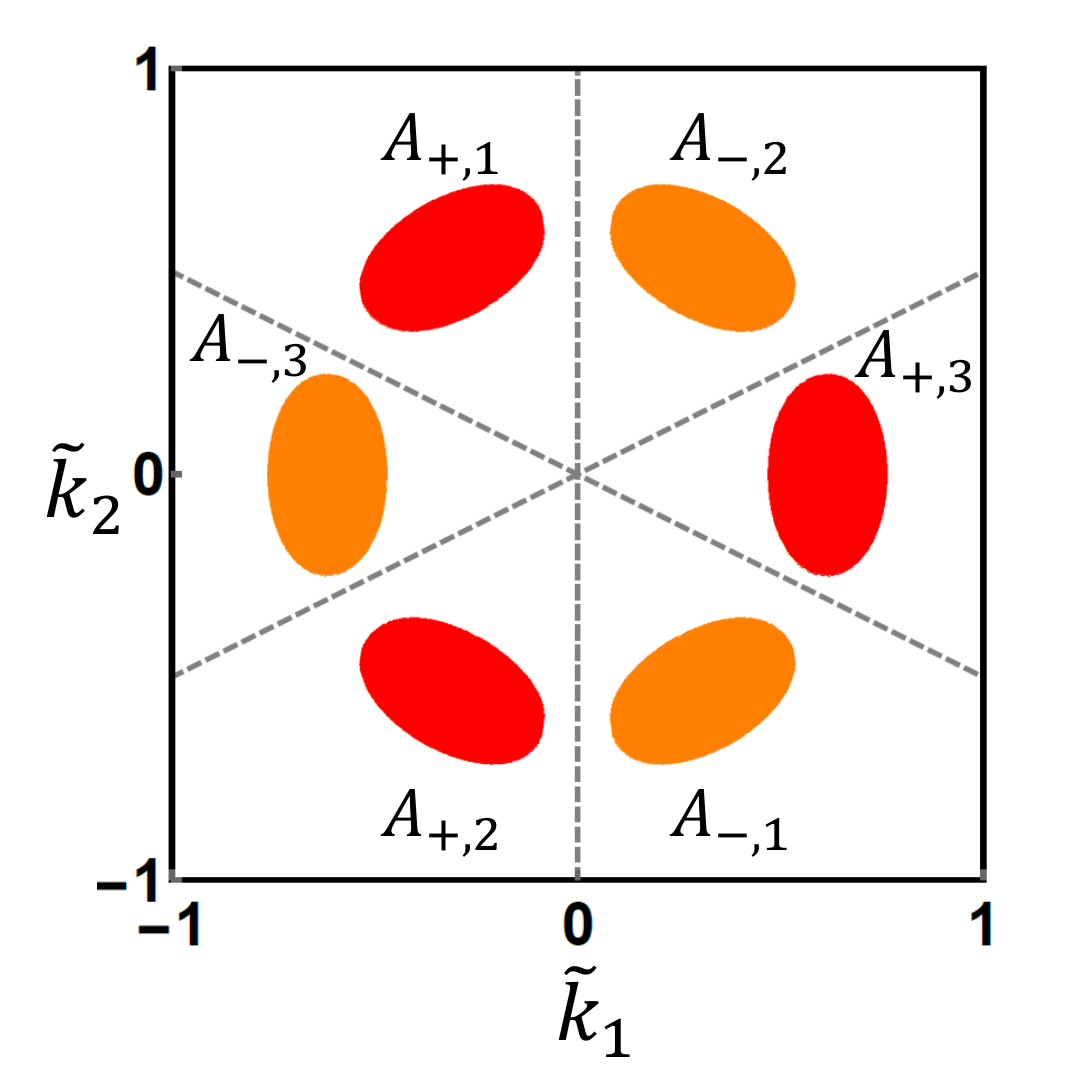}
    \caption{\label{fig:surf_MFB} This is the distribution of SMFBs for $|2 m| c_1=0.8$, $|2 m| c_2=0.5$, $\widetilde{\Delta}_0/|\mu|=1$ and $\widetilde{\Delta}_1/|\mu|=1.6$, where $\widetilde{\Delta}_0=\text{sgn}(c_1)\Delta_0$, $\widetilde{\Delta}_1=2m\mu a^2 \Delta_1$ and $\tilde{k}_{1,2}=k_{1,2}/\sqrt{2m\mu}$. The surface zero modes in red(orange) regions have $1$($-1$) chiral eigenvalue, and $A^{l_c,l_{\chi}}$'s are labeled according to the convention. The dashed lines are given by $k_{\shpa,1}=0$ and $k_{\shpa,2}=\pm k_{\shpa,1}/2$, where the surface zero modes cannot exist.
    }
\end{figure}

\section{Surface Majorana Flat Bands}
\label{sec:surf_MFB}
In this work, we choose $\mu<0$, $m<0$ and $c_1 c_2>0$, and focus on the case where $c_1\neq c_2$, $m_{\pm}<0$, and SMFBs exist on the $(111)$ surface.~\cite{yu2017Singlet-Quintetj=3/2SC}
To solve for SMFBs, we consider a semi-infinite configuration ($x_{\perp}<0$) of \eqnref{eq:H_BdG} along the $(111)$ direction with an open boundary condition at the $x_{\perp}=0$ surface, where $x_{\perp}$ labels the position along $(111)$.
In this case, the point group is reduced from $O_h$ to $C_{3v}$, which is generated by three-fold rotation $\hat{C}_3$ along the $(111)$ direction and the mirror operation $\hat{\Pi}$ perpendicular to the $(\bar{1}10)$ direction.
Although the translational invariance along $(111)$ is broken, the momentum $\bsl{k}_{\shpa}$ that lies inside the $(111)$ plane is still a good quantum number, and we define $k_{\shpa,1}$ and $k_{\shpa,2}$ along the $(11\bar{2})$ and $(\bar{1}10)$ directions, respectively.

Following \refcite{yu2017Singlet-Quintetj=3/2SC}, we find that SMFBs can exist in certain regions of the surface Brillouin zone, denoted as $A$ in \figref{fig:surf_MFB}, and originate from the non-trivial one-dimensional AIII bulk topological invariant $(N_{w}=\pm 2)$.
At each $\bsl{k}_{\shpa}\in A$, the semi-infinite model has two orthonormal solutions of zero energy that are localized near the $x_{\perp}=0$ surface and have the same chrial eigenvalues, coinciding with the bulk topological invariant $N_{w}=\pm 2$.

We label the creation operators for the two zero-energy solutions at $\bsl{k}_{\shpa}\in A$ as $b^{\dagger}_{i,\bsl{k}_{\shpa}}$ with $i=1,2$, and they satisfy the anti-commutation relation
\begin{equation}
\{ b^{\dagger}_{i,\bsl{k}_{\shpa}}, b_{j,\bsl{k}'_{\shpa}}\}=\delta_{ij}\delta_{\bsl{k}_{\shpa} \bsl{k}'_{\shpa}}\ .
\end{equation}
The subscript $i=1,2$ of $b^{\dagger}_{i,\bsl{k}_{\shpa}}$ can be regarded as the pseudospin index, since $b^{\dagger}_{i,\bsl{k}_{\shpa}}$ can furnish the same representation of TR, $\hat{C}_3$ and $\hat{\Pi}$ operators as a two dimensional $j=1/2$ fermion by choosing the convention
\begin{equation}
\label{eq:rep_TR_C3v_surf}
\left\{
\begin{array}{l}
\hat{\mathcal{T}}b^{\dagger}_{\bsl{k}_{\shpa}}\hat{\mathcal{T}}^{-1}=
b^{\dagger}_{-\bsl{k}_{\shpa}}\mathcal{T}_b \\
\hat{C}_3 b^{\dagger}_{\bsl{k}_{\shpa}}\hat{C}_3^{-1}=
b^{\dagger}_{C_3\bsl{k}_{\shpa}}C_{3,b} \\
\hat{\Pi} b^{\dagger}_{\bsl{k}_{\shpa}}\hat{\Pi}^{-1}=
b^{\dagger}_{\Pi\bsl{k}_{\shpa}}\Pi_{b}
\end{array}
\right. \ ,
\end{equation}
where $\mathcal{T}_b=i\sigma_2$, $C_{3,b}=e^{-i \sigma_3 \frac{\pi}{3}}$, $\Pi_b=-e^{-i \sigma_2 \frac{\pi}{2}}$, and $\sigma_{1,2,3}$ are Pauli matrices for the pseudospin of SMFBs.
Since the chiral matrix $\chi$ commutes with any operation in $C_{3v}$ and anti-commutes with TR operation, the chiral eigenvalue of $b^{\dagger}_{i,\bsl{k}_{\shpa}}$ is the same as $b^{\dagger}_{i,R\bsl{k}_{\shpa}}$, but opposite to $b^{\dagger}_{i,-\bsl{k}_{\shpa}}$, where $R\in C_{3v}$.
As a result, the surface zero-energy modes cannot exist on three lines parametrized by $k_{\shpa,1}=0$ and $k_{\shpa,2}=\pm k_{\shpa,1}/2$, dividing the region $A$ into six patches as shown in \figref{fig:surf_MFB}.
Since the chiral eigenvalues of the zero-energy modes in one patch are the same, we can label each patch as $A_{l_{\chi}, \l_c}$ with $l_{\chi}=\pm$ for the chiral eigenvalues $\pm 1$ and $l_c=1,2,3$ marking three patches related by $\hat{C}_{3}$ rotation.
Furthermore, we choose $A_{l_{\chi}, 3}$ to be symmetric under $k_{\shpa,2}\rightarrow -k_{\shpa,2}$, i.e. the mirror operation perpendicular to $(\bar{1}10)$.
Due to the PH symmetry, the surface zero modes at $\pm \bsl{k}_{\shpa}$ are related by
\begin{equation}
\label{eq:b_PH}
b^{\dagger}_{\bsl{k}_{\shpa}}(-\delta^{\chi}_{\bsl{k}_{\shpa}}\sigma_{2})=b^T_{-\bsl{k}_{\shpa}}\ ,
\end{equation}
where $\delta^{\chi}_{\bsl{k}_{\shpa}}$ is the chiral eigenvalue of the zero modes at $\bsl{k}_\shpa$, i.e. $\delta^{\chi}_{\bsl{k}_{\shpa}}=\pm 1$ for $\bsl{k}_{\shpa}\in A_{\pm}$ with $A_{l_\chi}=\cup_{l_c} A_{l_\chi,l_c}$. 
TR and $C_{3v}$ symmetries imply $\delta^{\chi}_{-\bsl{k}_{\shpa}}=-\delta^{\chi}_{\bsl{k}_{\shpa}}$ and $\delta^{\chi}_{R\bsl{k}_{\shpa}}=\delta^{\chi}_{\bsl{k}_{\shpa}}$ with $R\in C_{3v}$.
(See Appendix.\ref{app:surf_modes} for details.)


\section{Mean-field Order Parameters of Surface Majorana Flat Bands}
\label{sec:MF_order_MFB}
Due to the divergent DOS, the interaction may result in the nonvanishing order parameters at the surface and give rise to a gap of SMFBs.
In this section, we study the possible mean-field order parameters on the $(111)$ surface that preserve the in-plane translation symmetry.
We find that the order parameters must break the TR symmetry in order to gap out the SMFB; all the TR-breaking surface order parameters are classified based on the IRs of $C_{3v}$ and their physical origins are identified.
Then, to the leading order approximation where the surface order parameters are independent of $\bsl{k}_{\shpa}$ in each of the surface mode regions, we find the SMFBs can be generally gapped out by these order parameters, and the gapless modes are only possible for certain IRs with certain finely tuned values of parameters.
We further study the LDOS structure of SMFBs in the presence of various order parameters and find the splitting patterns of the LDOS peak can be used to distinguish different order parameters as summarized in \figref{fig:LDOS_no_imp} and \ref{fig:LDOS_peak_split}.

\subsection{Symmetry Classification and Physical Origin}

The general form of translationally invariant fermion-bilinear terms for SMFBs can be constructed as
\begin{equation}
\label{eq:H_mf}
H_{mf}=\frac{1}{2}\sum_{\bsl{k}_{\shpa}\in A}  b^{\dagger}_{\bsl{k}_{\shpa}} m(\bsl{k}_{\shpa}) b_{\bsl{k}_{\shpa}}+const.\ ,
\end{equation}
where $m(\bsl{k}_{\shpa})$ is a $2 \times 2$ Hermitian matrix.
The PH symmetry makes $m(\bsl{k}_{\shpa})$ satisfy $m(\bsl{k}_{\shpa})=-\sigma_2\  m^T(-\bsl{k}_{\shpa})\sigma_2$ up to a shift of ground state energy based on \eqnref{eq:b_PH}, while TR symmetry requires $\mathcal{T}_b m^*(-\bsl{k}_{\shpa})\mathcal{T}_b^{\dagger}= m(\bsl{k}_{\shpa})$ according to \eqnref{eq:rep_TR_C3v_surf}.
As a result, the combination of PH and TR symmetries, which is equivalent to the chiral symmetry, leads to $m(\bsl{k}_{\shpa})=0$, indicating that the existence of a non-vanishing fermion bilinear term $m(\bsl{k}_{\shpa})$ for the SMFBs requires the breaking of TR symmetry, i.e.
\eq{
\mathcal{T}_b m^*(-\bsl{k}_{\shpa})\mathcal{T}_b^{\dagger}=-m(\bsl{k}_{\shpa})\ .
}
As the $C_{3v}$ point group symmetry can also be spontaneously broken by these fermion-bilinear terms, we can further classify these TR-breaking order parameters according to the IR of $C_{3v}$, of which the character table (\tabref{tab:cha_C3v}) is shown in Appendix.\ref{app:conv_expn}.
Since $C_{3v}$ has three IRs $A_1$, $A_2$ and $E$, \eqnref{eq:H_mf} can be expressed as the linear combination of the three corresponding parts
\begin{equation}
\label{eq:H_mf_IR}
m(\bsl{k}_{\shpa})=m_{A_1}(\bsl{k}_{\shpa})+m_{A_2}(\bsl{k}_{\shpa})+m_{E}(\bsl{k}_{\shpa})\ .
\end{equation}
Here the $A_1$ term $m_{A_1}(\bsl{k}_{\shpa})$ preserves $C_{3v}$ symmetry, and
the $A_2$ term $m_{A_2}(\bsl{k}_{\shpa})$ preserves $\hat{C}_{3}$ symmetry but has odd mirror parity.
The $E$ term has the expression $m_{E}(\bsl{k}_{\shpa})=a_1 m_{E,1}(\bsl{k}_{\shpa})+a_2 m_{E,2}(\bsl{k}_{\shpa})$ with $(m_{E,1}(\bsl{k}_{\shpa}), m_{E,2}(\bsl{k}_{\shpa}))$ a two-component vector that can furnish a $E$ IR; it breaks the entire $C_{3v}$ symmetry except for some special values of $(a_1,a_2)$, e.g. one of the three mirrors is preserved but the $\hat{C}_3$ is broken for $(a_1,a_2)\propto (1,0), (1,\sqrt{3})$ or $(1,-\sqrt{3})$.

Next we illustrate the physical origin of each term in \eqnref{eq:H_mf_IR} by considering the following on-site mean-field Hamiltonian that are independent of $\bsl{k}_{\shpa}$
\begin{eqnarray}
\label{eq:Ht_mf}
&&\widetilde{H}_{mf}=\sum_{\bsl{k}_{\shpa}}^{A}\int_{-\infty}^{0} d x_{\perp}  [c^{\dagger}_{\bsl{k}_{\shpa},x_{\perp}} \widetilde{M}(x_{\perp}) c_{\bsl{k}_{\shpa},x_{\perp}}+\\
&& \frac{1}{2} c^{\dagger}_{\bsl{k}_{\shpa},x_{\perp}} \widetilde{D}(x_{\perp}) (c^{\dagger}_{-\bsl{k}_{\shpa},x_{\perp}})^T + \frac{1}{2} c^T_{-\bsl{k}_{\shpa},x_{\perp}} \widetilde{D}^{\dagger}(x_{\perp}) c_{\bsl{k}_{\shpa},x_{\perp}}]\ ,\nonumber
\end{eqnarray}
where $\widetilde{M}^{\dagger}(x_{\perp})=\widetilde{M}(x_{\perp})$ and $-\widetilde{D}^T(x_{\perp})=\widetilde{D}(x_{\perp})$.
\eqnref{eq:H_mf} can be obtained by projecting the above Hamiltonian onto the surface, and such projection does not change the symmetry properties.
Since $m(\bsl{k}_{\shpa})$ must be TR odd in order to be non-vanishing, it requires $\widetilde{M}(x_{\perp})$ and $\widetilde{D}(x_{\perp})$ to be TR odd.
Then, the TR-breaking $\widetilde{M}$ and $\widetilde{D}$ can be classified into different IRs of $C_{3v}$:
\begin{equation}
\widetilde{M}(x_{\perp})=\widetilde{M}_{A_1}(x_{\perp})+\widetilde{M}_{A_2}(x_{\perp})+\widetilde{M}_E(x_{\perp})\ ,
\end{equation}
and
\begin{equation}
\widetilde{D}(x_{\perp})=\widetilde{D}_{A_1}(x_{\perp})+\widetilde{D}_{A_2}(x_{\perp})+\widetilde{D}_E(x_{\perp})\ ,
\end{equation}
where $\widetilde{M}_{\beta}(x_{\perp})$ and $\widetilde{D}_{\beta}(x_{\perp})$ can only give rise to $m_{\beta}(\bsl{k}_{\shpa})$ in \eqnref{eq:H_mf_IR} with $\beta=A_1, A_2, E$.
(See Appendix.\ref{app:H_mf_c2b} for details.)
Concretely, we have
\begin{equation}
\label{eq:Mt_Dt}
\left\{
\begin{array}{l}
\widetilde{M}_{A_1}(x_{\perp})=\zeta_2(x_{\perp}) n_2\\
\widetilde{M}_{A_2}(x_{\perp})=\sum_{j=3}^5 \zeta_j(x_{\perp}) n_j\\
\widetilde{M}_{E}(x_{\perp})=\sum_{j=8}^{10} \bsl{\zeta}_j(x_{\perp})\cdot \bsl{n}_j\\
\widetilde{D}_{A_1}(x_{\perp})=\sum_{j=0}^1 i \zeta_j(x_{\perp}) n_j \gamma\\
\widetilde{D}_{A_2}(x_{\perp})=0\\
\widetilde{D}_{E}(x_{\perp})=\sum_{j=6}^7 i \bsl{\zeta}_j(x_{\perp})\cdot \bsl{n}_j \gamma
\end{array}
\right.\ ,
\end{equation}
where $n_i$'s are listed in \tabref{tab:n_class} of Appendix.\ref{app:conv_expn}, and $\zeta_j(x_{\perp})$'s are real.
Physically, $n_0\gamma$ corresponds to the singlet pairing, $n_1\gamma$, $\bsl{n}_6\gamma$ and $\bsl{n}_7\gamma$ generate quintet pairings, and $n_4, n_{8,1}, n_{8,2}$ give FM in $(111)$, $(1\bar{1}0)$ and $(11\bar{2})$ directions, respectively.
Since $n_2, n_3, n_5, \bsl{n}_9$ and $\bsl{n}_{10}$ can be represented by the linear combinations of $c^{\dagger}_{\bsl{k}_{\shpa},x_{\perp}}S^{3m} c_{\bsl{k}_{\shpa},x_{\perp}}$ with the septet spin tensor $S^{3m}$($m=-3,-2,...,3$),
we dub these terms the spin-septet order parameters.
As a summary, $m_{A_1}(\bsl{k}_{\shpa})$ can be generated by the singlet pairing, the quintet pairing, and the spin-septet order parameter; $m_{A_2}(\bsl{k}_{\shpa})$ can be generated by $(111)$-directional ferromagnetism (FM) and the spin-septet order parameter; $m_{E}(\bsl{k}_{\shpa})$ can be generated by the quintet pairing, the FM perpendicular to the $(111)$ direction, and the spin-septet order parameter.

\begin{table}
  \centering
  \begin{tabular}{c|m{5 cm}|c|c|c}
    \hline
    $C_{3v}$ & Bases & TR & PH & $\chi$\\ \hline
    $A_1$ & $\sum_{l_{\chi},l_c} \delta^{l_{\chi}l_c}_{\bsl{k}_{\shpa}}=1$ for $\bsl{k}_{\shpa}\in A$ & + & $+$ & $+$ \\
    $A_1$ & $\delta^{\chi}_{\bsl{k}_{\shpa}}=\sum_{l_{\chi},l_c} l_{\chi} \delta^{l_{\chi}l_c}_{\bsl{k}_{\shpa}}$ & $-$ & $-$ & $+$\\
    $E$ &  $(\delta^{E_1,+}_{\bsl{k}_{\shpa}},\delta^{E_2,+}_{\bsl{k}_{\shpa}})$  & + & $+$ & $+$\\
    $E$ &  $(\delta^{E_1,-}_{\bsl{k}_{\shpa}},\delta^{E_2,-}_{\bsl{k}_{\shpa}})$  & $-$& $-$ & $+$\\ \hline
    $A_1$ & $\sigma_0$ & $+$ & $+$ & $+$\\
    $A_2$ & $\sigma_3$ & $-$ & $-$ & $+$\\
    $E$ &  $(-\sigma_2,\sigma_1)$  & $-$& $-$ & $+$\\ \hline
    $A_1$ & $\rho_0$ & $+$ & $+$ & $+$\\
    $A_1$ & $\rho_1$ & $+$ & $-$ & $-$\\
    $A_1$ & $\rho_2$ & $+$ & $-$ & $-$\\
    $A_1$ & $\rho_3$ & $-$ & $-$ & $+$\\ \hline
    $A_1$ & $\Lambda_1=\lambda_0$ & $+$ & $+$ & $+$\\
    $A_1$ & $\Lambda_2=\frac{1}{\sqrt{2}}(\lambda_1+\lambda_4+\lambda_6)$ & $+$ & $+$ & $+$\\
    $A_2$ & $\Lambda_3=\frac{1}{\sqrt{2}}(\lambda_2-\lambda_5+\lambda_7)$ & $-$ & $-$ & $+$\\
    $E$ &  $\bsl{\Lambda}_4=\frac{\sqrt{3}}{2}(\lambda_8,-\lambda_3)$  & $+$& $+$ & $+$\\
    $E$ &  $\bsl{\Lambda}_5=\sqrt{\frac{3}{8}}(\lambda_5+\lambda_7,\frac{-2\lambda_2-\lambda_5+\lambda_7}{\sqrt{3}})$  & $-$& $-$ & $+$\\
    $E$ &  $\bsl{\Lambda}_6=\sqrt{\frac{3}{8}}(\frac{-2\lambda_1+\lambda_4+\lambda_6}{\sqrt{3}},\lambda_4-\lambda_6)$  & $+$& $+$ & $+$\\ \hline
    \end{tabular}
  \caption{ The irreducible representations of $C_{3v}$  generated by $\delta^{l_{\chi}l_c}_{\bsl{k}_{\shpa}}$, $\sigma_l$, $\rho_l$ or $\lambda_l$ with their parities under TR, PH and chiral operation.
  The transformation of $\delta^{l_{\chi}l_c}_{\bsl{k}_{\shpa}}$ is defined as $\delta^{l_{\chi}l_c}_{\bsl{k}_{\shpa}}\rightarrow \delta^{l_{\chi}l_c}_{R^{-1}\bsl{k}_{\shpa}}$, the transformation of $\sigma_l$ is $\sigma_l\rightarrow R_b \sigma_l R_b^{\dagger}$, the transformation of $\rho_l$ is $\rho_l \rightarrow R_\chi \rho_l R_\chi^{\dagger}$ and the transformation of $\lambda_l$ is $\lambda_l \rightarrow R_c \lambda_l R_c^{\dagger}$, where $R=-1,-1,1,C_3, \Pi$,  $R_b=i\sigma_2 K ,\sigma_2 K ,\mathds{1},C_{3,b}, \Pi_b$,  $R_\chi=\mathcal{T}_\chi K ,\mathcal{C}_\chi K ,\chi_\chi,C_{3,\chi}, \Pi_\chi$ and $R_c=\mathcal{T}_c K ,\mathcal{C}_c K,\chi_c,C_{3,c}, \Pi_c$ for TR, PH, $
  \chi$, $C_3$ and $\Pi$, respectively, and $K$ is the complex conjugate operation.
  The parity $\alpha=\pm$ is defined as $X\rightarrow \alpha X$ under the operation of TR, PH or $\chi$ and thus being TR, PH and $\chi$ symmetric correspond to $\alpha=+,-,-$, respectively.
  $l_{\chi}=\pm$, $l_{c}=1,2,3$ and $\delta^{l_{\chi}l_c}_{\bsl{k}_{\shpa}}$ is equal to $1$ if $\bsl{k}_{\shpa}\in A_{l_{\chi},l_c}$ and $0$ otherwise. $(\delta^{E_1,+}_{\bsl{k}_{\shpa}},\delta^{E_2,+}_{\bsl{k}_{\shpa}})=\left(\sum_{l_{\chi}}\frac{1}{2} (\delta^{l_{\chi},1}_{\bsl{k}_{\shpa}}+\delta^{l_{\chi},2}_{\bsl{k}_{\shpa}}-2 \delta^{l_{\chi},3}_{\bsl{k}_{\shpa}})\right.$ , $\left. \sum_{l_{\chi}}\frac{\sqrt{3}}{2} (-\delta^{l_{\chi},1}_{\bsl{k}_{\shpa}}+\delta^{l_{\chi},2}_{\bsl{k}_{\shpa}})\right)$ and $(\delta^{E_1,-}_{\bsl{k}_{\shpa}},\delta^{E_2,-}_{\bsl{k}_{\shpa}})=\left(\sum_{l_{\chi}}\frac{l_{\chi}}{2} (\delta^{l_{\chi},1}_{\bsl{k}_{\shpa}}+\delta^{l_{\chi},2}_{\bsl{k}_{\shpa}}-2 \delta^{l_{\chi},3}_{\bsl{k}_{\shpa}})\right.$ , $\left. \sum_{l_{\chi}}\frac{l_{\chi}\sqrt{3}}{2} (-\delta^{l_{\chi},1}_{\bsl{k}_{\shpa}}+\delta^{l_{\chi},2}_{\bsl{k}_{\shpa}})\right)$.} \label{tab:IR_C3v_TR_PH_chi}
\end{table}

\subsection{Surface Local Density of States}

In the following, we focus on the order parameters that are independent of $\bsl{k}_{\shpa}$ in every one of six surface mode regions $A_{l_{\chi}, l_c}$'s.
In this case, \eqnref{eq:H_mf_IR} can be expanded as
\begin{equation}
\label{eq:m_unif_gen}
m(\bsl{k}_{\shpa})=\sum_{l=0}^4\sum_{l_{\chi}=\pm}\sum_{l_c=1}^3 f_{l}^{l_{\chi} l_c}\sigma_l \delta_{\bsl{k}_{\shpa}}^{l_{\chi} l_c}\ ,
\end{equation}
where $f_{l}^{l_{\chi} l_c}$ is real, $\delta^{l_{\chi}l_c}_{\bsl{k}_{\shpa}}=1$ if $\bsl{k}_{\shpa}\in A_{l_{\chi},l_c}$ and $\delta^{l_{\chi}l_c}_{\bsl{k}_{\shpa}}=0$ otherwise, and $\sigma_l$ labels the Pauli matrix for pseudospin.
Then, for any symmetry transformation of $m(\bsl{k})$, we can convert the transformation of pseudospin index and $\bsl{k}_{\shpa}$ dependence of $m(\bsl{k})$ to the transformation of $\sigma_l$ and $\delta^{l_{\chi}l_c}_{\bsl{k}_{\shpa}}$, respectively.
Based on the symmetry transformation, we can classify $\delta^{l_{\chi}l_c}_{\bsl{k}_{\shpa}}$ and $\sigma_l$ according to the IRs of $C_{3v}$ and the parities under TR, PH and $\chi$, as shown in the top and second top parts of \tabref{tab:IR_C3v_TR_PH_chi}, respectively.
The symmetry classification of TR-odd terms in $m(\bsl{k}_{\shpa})$ can be obtained by the tensor product of $\sigma_l$ and $\delta^{l_{\chi}l_c}_{\bsl{k}_{\shpa}}$, as shown in \tabref{tab:N} of Appendix.\ref{app:conv_expn} with various terms labeled by $N_i$'s.
As a result, we have the following general expressions of the order parameters in different IRs of $C_{3v}$:
\begin{equation}
\label{eq:H_mf_A1}
m_{A_1}(\bsl{k}_{\shpa})=\sum_{j=1}^2 m_j N_{j}(\bsl{k}_{\shpa})\ ,
\end{equation}
\begin{equation}
\label{eq:H_mf_A2}
m_{A_2}(\bsl{k}_{\shpa})=\sum_{j=3}^4 m_j N_{j}(\bsl{k}_{\shpa})\ ,
\end{equation}
and
\begin{equation}
\label{eq:H_mf_E}
m_{E}(\bsl{k}_{\shpa})=\sum_{j=5}^8 \bsl{m}_j\cdot \bsl{N}_{j}(\bsl{k}_{\shpa})\ .
\end{equation}
Here all $m_j$'s are real.

With \eqnref{eq:H_mf_A1}-\eqref{eq:H_mf_E}, we next discuss the energy spectrum and LDOS of SMFBs after including these order parameters.
Due to the PH symmetry, only half of the energy spectrum (non-negative energy part) gives the quasi-particle LDOS of SMFBs.
However, it is more convenient to study the full spectrum, since the LDOS, which is probed by the tunneling conductance of STM, must symmetrically distribute with respect to the zero energy in experiments~\cite{tinkham1996introductionSC}.
%
%
Since the order parameters in each patch are $\bsl{k}_{\shpa}$-independent, we choose the mode at the geometric center $\bsl{K}^{l_{\chi}, l_c}_{\shpa}$ of each patch $A_{l_{\chi},l_c}$ as the representative mode.
In the following, we only consider the representative modes and use the term ``degeneracy" to refer to the {\it extra} degeneracy determined by the symmetry, excluding the large degeneracy given by the flatness of the dispersion in each patch.
For convenience, we define the creation operator $b^{\dagger}_{i,l_{\chi},l_{c}}=b^{\dagger}_{i,\bsl{K}^{l_{\chi}, l_c}_{\shpa}}$ to label the representative mode in the patch $A_{l_{\chi},l_c}$ with the pseudo-spin index $i$.
Since only the uniform order parameters are considered,
$l_{\chi}$ and $l_c$ are good quantum numbers, while different pseudo-spin components (the $\sigma_l$ part) are typically coupled by the order parameter $m(\bsl{k}_{\shpa})$.
Thus, we introduce the band index $s=\pm$ and label the eigen-mode as $\widetilde{b}^{\dagger}_{s,l_\chi,l_c}=\sum_{i}X^{s,l_\chi,l_c}_i b^{\dagger}_{i,l_{\chi},l_{c}}$ with
\eq{
\sum_{l_\chi,l_c}m(\bsl{K}^{l_{\chi}, l_c}_{\shpa})X^{s,l_\chi,l_c}=\sum_{l_\chi,l_c} E^{s,l_\chi,l_c} X^{s,l_\chi,l_c}
}
the eigen-equation.

\begin{figure}[t]
    \centering
    \includegraphics[width=\columnwidth]{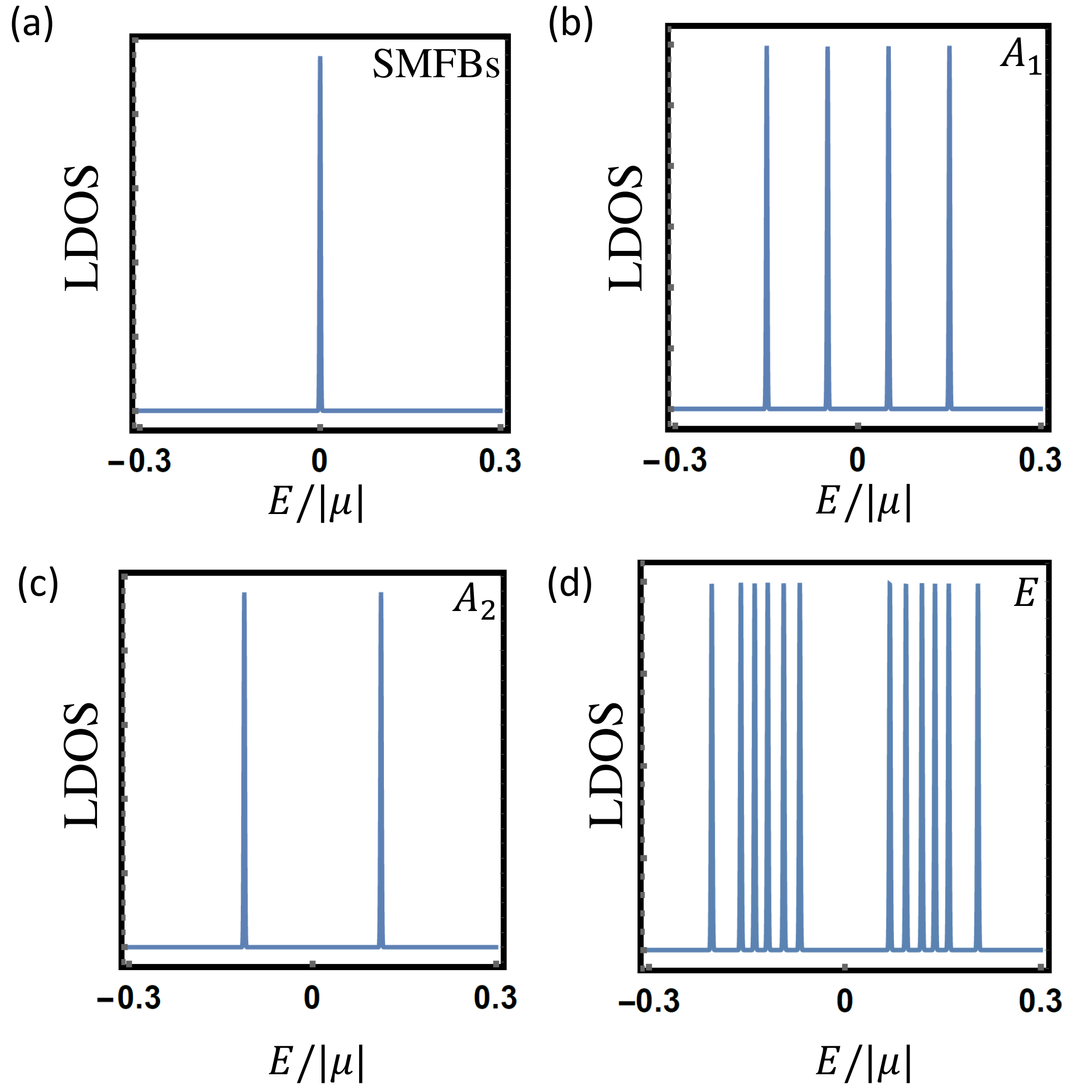}
    \caption{\label{fig:LDOS_no_imp} (a), (b), (c) and (d) show the LDOS on the $(111)$ surface as a function of the energy ($E/|\mu|$) without any order parameters, with  the $A_1$ order parameter, with the $A_2$ order parameter and with the $E$ order parameter, respectively.
Due to PH symmetry, only non-negative-energy half of the LDOS is physical.
The broadening of each peak is plotted via Gaussian distribution with standard deviation being $10^{-3}$.The parameters choices for each order if exist are $m_1/|\mu|=0.05$ and $m_2/|\mu|=0.1$ for the $A_1$ order parameter (\ref{eq:H_mf_A1}), $m_3/|\mu|=0.05$ and $m_4/|\mu|=-0.1$ for the $A_2$ order parameter (\ref{eq:H_mf_A2}), and $\bsl{m}_5/|\mu|=(0.01,0.02)$,$\bsl{m}_6/|\mu|=(0.03,0.04)$,$\bsl{m}_7/|\mu|=(0.05,0.06)$ and $\bsl{m}_{8}/|\mu|=(0.07, 0.08)$ for the $E$ order parameter (\ref{eq:H_mf_E}).
Here we don't show the numbers on the vertical axis~\cite{Bi2019TBG} since only the position of LDOS peak can be probed in the STM experiments.
}
\end{figure}

Without any order parameters, all these 12 modes, including 6 patches and 2 pseudospin components, are degenerate and thus the SMFBs has a zero-bias peak for LDOS, as shown in \figref{fig:LDOS_no_imp}a.
For the $A_1$ order $m_{A_1}(\bsl{k}_{\shpa})$, the eigen-energies are given by $m_1 \delta^{\chi}_{\bsl{k}_{\shpa}}\pm |m_2|$, and once $|m_1|\neq|m_2|$, all the zero energy peaks will be split for SMFBs.
As a result, the LDOS of the $A_1$ order parameter typically has 4 peaks shown in \figref{fig:LDOS_no_imp}b.
This peak structure of LDOS can be understood from symmetry consideration.
Due to the breaking of TR symmetry, as well as the chiral symmetry, we only need to consider the the point group symmetry $C_{3v}$.
As mentioned before, any operation in $C_{3v}$ does not change the $l_{\chi}$ index, and since $A_{1}$ order parameter is $C_{3v}$ invariant, the band index $s$ cannot be changed either.
The $C_3$ rotation only transforms the $l_c={1,2,3}$ index counter-clockwise, resulting in the three-fold degeneracy among the eigen-modes $\widetilde{b}^{\dagger}_{s,l_{\chi},l_c}$ with the same $s$ and $l_{\chi}$.
%
On the other hand, $\Pi$ interchanges $l_c=1,2$ and makes sure $\widetilde{b}^{\dagger}_{s,l_{\chi},1}$ has the same energy as $\widetilde{b}^{\dagger}_{s,l_{\chi},2}$, meaning that $\Pi$ does not give extra constraints compared with $C_3$.
Thus, there are $12/3=4$ peaks in the LDOS of the $A_1$ order parameter with each peak of 3-fold degeneracy.
For the $A_2$ order parameter $m_{A_2}(\bsl{k}_{\shpa})$, the eigen-energies are given by $\pm \sqrt{m_3^2+m_4^2}$, leading to 2 peaks in the LDOS (\figref{fig:LDOS_no_imp}c), resulted from the six-fold degeneracy of each eigen-energy due to the symmetry.
Among the six-fold degeneracy, three-fold degeneracy is due to translational invariance and $C_3$ symmetry as the $A_1$ order parameter, meaning that $\widetilde{b}^{\dagger}_{s,l_\chi,1}$, $\widetilde{b}^{\dagger}_{s,l_\chi,2}$ and $\widetilde{b}^{\dagger}_{s,l_\chi,3}$ have the same energy.
The remaining double degeneracy originates from the combination of the odd mirror parity of the $A_2$ order parameter and the PH symmetry, i.e.
$\Pi_b \sigma_2 m_{A_2}^* (-\Pi^{-1}\bsl{k}_{\shpa}) (\Pi_b \sigma_2)^{\dagger}=m_{A_2}(\bsl{k}_{\shpa})$.
This combined symmetry does not change the band index $s$, but transforms $l_\chi$ as $+\leftrightarrow -$ and $l_c$ as $1\leftrightarrow 2$.
As a result, $\widetilde{b}^{\dagger}_{s,\pm,l_c}$ with fixed $s$ and $l_c$ also have the same energy, giving the extra double degeneracy.
For the $E$ order parameter $m_{E}(\bsl{k}_{\shpa})$, the eigen-energies are $\sum_{l_\chi,l_c} (l_\chi \bar{m}_{l_c}\pm \bar{m}'_{l_c})\delta^{l_{\chi},l_c}_{\bsl{k}_{\shpa}}$, where
\eqa{
&\bar{m}_{1}=\frac{m_{5,1}}{2}-\frac{\sqrt{3}}{2}m_{5,2}\\
&\bar{m}_{2}=\frac{m_{5,1}}{2}+\frac{\sqrt{3}}{2}m_{5,2}\ ,\ \bar{m}_{3}=-m_{5,1}\\
&\bar{m}'_{1}=[(\frac{\sqrt{3}}{2}m_{6,1}+\frac{m_{6,2}}{2})^2+(-m_{7,1}+\frac{m_{8,1}}{2}+\frac{\sqrt{3}}{2}m_{8,2})^2\\
&+(m_{7,2}-\frac{\sqrt{3}}{2}m_{8,1}+\frac{m_{8,2}}{2})^2]^{1/2}\\
&\bar{m}'_{2}=[(-\frac{\sqrt{3}}{2}m_{6,1}+\frac{m_{6,2}}{2})^2+(-m_{7,1}+\frac{m_{8,1}}{2}-\frac{\sqrt{3}}{2}m_{8,2})^2\\
&+(m_{7,2}+\frac{\sqrt{3}}{2}m_{8,1}+\frac{m_{8,2}}{2})^2]^{1/2}\\
&\bar{m}'_{3}=[m^2_{6,2}+(m_{7,1}+m_{8,1})^2+(m_{7,2}-m_{8,2})^2]^{1/2}\ .
}
Therefore, all the modes are typically split for the $E$ order and the corresponding LDOS generally has 12 peaks shown in \figref{fig:LDOS_no_imp}d.

We would like to mention that if including the momentum dependence of the surface order parameter in each surface-mode region, it can broaden the LDOS peaks in \figref{fig:LDOS_no_imp}.
In addition, the momentum dependence may also lead to the existence of arcs of surface zero modes in certain small parameter regions as discussed Appendix.\ref{app:MZM_arc}.

\section{Impurity Effect}
\label{sec:imp_MFB}
In this section, we will study the effect of surface non-magnetic and magnetic impurities.
The effect of non-magnetic impurity on SMFBs in the absence of the mean-field order parameters has been studied in \refcite{Sato2011TopoABS,Ikegaya2015APE,Ikegaya2017MZMDirtyNSC,Ikegaya2018SymABSDirty}, showing that any non-magnetic impurity can generally induce a local gap for the SMFBs of DIII TNLS.
Our work here aims to present a systematic study on how the LDOS of SMFBs is split around a single non-magnetic or magnetic impurity in the absence/presence of the mean-field order parameters.

\subsection{Preliminaries}

To consider the local potential, we first need to transform SMFBs to the real space with
\begin{equation}
\label{eq:b_r_k}
b^{\dagger}_{l_{\chi},l_c,i,\bsl{r}_{\shpa}}=\frac{1}{\sqrt{\mathcal{S}_{\shpa}}}\sum_{\bsl{k}_{\shpa}}^{A_{l_{\chi},l_c}}e^{-i\bsl{k}_{\shpa}\cdot \bsl{r}_{\shpa}}b^{\dagger}_{i,\bsl{k}_{\shpa}}\ ,
\end{equation}
where the momentum summation is limited into the surface mode region $A_{l_{\chi},l_c}$.
Under the symmetry operations, the indexes $i,l_\chi,l_c$ of  $b^{\dagger}_{l_{\chi},l_c,i,\bsl{r}_{\shpa}}$ defined here are transformed in the same way as those of $b^{\dagger}_{i,l_{\chi},l_c}$ defined in \secref{sec:MF_order_MFB}.
In the following, we adopt the following approximation
\begin{equation}
\label{eq:appro_local_d}
\frac{1}{S_{\shpa}}\sum_{\bsl{k}_{\shpa}}^{A_{l_{\chi},l_{c}}}e^{i(\bsl{k}_{\shpa}-\bsl{K}^{l_{\chi},l_{c}}_{\shpa})\cdot \bsl{r}_{\shpa}}\approx \delta^{(2)}(\bsl{r}_{\shpa})\ ,
\end{equation}
resulting in
\begin{equation}
\{ b^{\dagger}_{l_{\chi},l_c,i,\bsl{r}_{\shpa}}, b_{l_{\chi}',l_c',i',\bsl{r}_{\shpa}'}\}=\delta_{l_{\chi}l_{\chi}'}\delta_{l_c l_c'}\delta_{i i'}\delta^{(2)}(\bsl{r}_{\shpa}-\bsl{r}_{\shpa}')\ .
\end{equation}
Further, we define
\eq{
d^{\dagger}_{\bsl{r}_{\shpa}}=(b^{\dagger}_{{+,1},\bsl{r}_{\shpa}}, b^{\dagger}_{{+,2},\bsl{r}_{\shpa}}, b^{\dagger}_{{+,3},\bsl{r}_{\shpa}}, b^{\dagger}_{{-,1},\bsl{r}_{\shpa}}, b^{\dagger}_{{-,2},\bsl{r}_{\shpa}}, b^{\dagger}_{{-,3},\bsl{r}_{\shpa}})
}
for convenience.

The behavior of $d^{\dagger}_{\bsl{r}_{\shpa}}$ under the symmetry transformation is crucial for the understanding of LDOS.
In general, the relation required by the PH symmetry has the form $d^{\dagger}_{\bsl{r}_{\shpa}}\mathcal{C}_d=d_{\bsl{r}_{\shpa}}^T$, and the transformation under TR, $\hat{C}_3$, and $\hat{\Pi}$ operations reads $\hat{\mathcal{T}}d^{\dagger}_{\bsl{r}_{\shpa}}\hat{\mathcal{T}}^{-1}=
d^{\dagger}_{\bsl{r}_{\shpa}}\mathcal{T}_d$,
$\hat{C}_3 d^{\dagger}_{\bsl{r}_{\shpa}}\hat{C}_3^{-1}=
d^{\dagger}_{C_3\bsl{r}_{\shpa}}C_{3,d}$, and
$\hat{\Pi} d^{\dagger}_{\bsl{r}_{\shpa}}\hat{\Pi}^{-1}=
d^{\dagger}_{\Pi\bsl{r}_{\shpa}}\Pi_{d}$, respectively.
As $d^{\dagger}_{\bsl{r}_{\shpa}}$, besides $\bsl{r}_\shpa$, carries three indexes $l_\chi,l_c,i$ that transform independently under the symmetry operation,  the transformation matrices presented above should be in the tensor product form as
\eqa{
\label{eq:sym_d}
&\mathcal{C}_d=\mathcal{C}_{\chi}\otimes \mathcal{C}_c \otimes \sigma_2\ \text{with}\ \mathcal{C}_{\chi}=-i\rho_2\ \text{and}\ \mathcal{C}_c=\lambda_0\ ,\\
&\mathcal{T}_d=\mathcal{T}_{\chi}\otimes \mathcal{T}_{c}\otimes \mathcal{T}_b\ \text{with}\ \mathcal{T}_{\chi}=\rho_1\ \text{and}\ \mathcal{T}_{c}=\lambda_0\ ,\\
&C_{3,d}=C_{3,\chi}\otimes C_{3,c} \otimes C_{3,b}\ \text{with}\ C_{3,\chi}=\rho_0\ ,\\
&\Pi_d=\Pi_{\chi}\otimes \Pi_{c}\otimes \Pi_{b}\ \text{with}\ \Pi_{\chi}=\rho_0\ ,
}
where $C_{3,c}=\exp(-i \frac{\lambda_2-\lambda_5+\lambda_7}{\sqrt{3}} \frac{2\pi}{3})$, $\Pi_c=-\exp(i \frac{\lambda_5+\lambda_7}{\sqrt{2}} \pi)$, $\rho_i$'s are Pauli matrices for $l_\chi=\pm$ index, $\sigma_i$'s are for the pseudo-spin of the surface modes as before, and $\lambda_i$'s are Gell-Mann matrices (Appendix.\ref{app:conv_expn}) for $l_c=1,2,3$ index with $\lambda_0$ the $3\times 3$ identity matrix.
%
%
In addition, the representation of the translation operator perpendicular to $(111)$ direction is
$\hat{T}_{\bsl{x}_{\shpa}}d^{\dagger}_{\bsl{r}_{\shpa}} \hat{T}^{-1}_{\bsl{x}_{\shpa}}=d^{\dagger}_{\bsl{r}_{\shpa}+\bsl{x}_{\shpa}}$.

With the above definition of $d^{\dagger}_{\bsl{r}_{\shpa}}$ operator, we next consider the Hamiltonian that describes the effect of a surface impurity on the SMFBs, given by
\begin{equation}
\label{eq:H_imp_d}
H_V=\int d^2 r_{\shpa} d_{\bsl{r}_{\shpa}}^{\dagger}M_V(\bsl{r}_{\shpa})d_{\bsl{r}_{\shpa}}+const.\ ,
\end{equation}
where $M_{V}(\bsl{r}_{\shpa})$ is Hermitian, the PH symmetry requires $\mathcal{C}_d M_V^*(\bsl{r}_{\shpa}) \mathcal{C}_d^{\dagger}=-M_V(\bsl{r}_{\shpa})$, and the impurity is chosen to be at $\bsl{r}_{\shpa}=0$ without the loss of generality.
Such form of impurity Hamiltonian is justified in Appendix.\ref{app:H_d_bases}.
$M_V(\bsl{r}_{\shpa})$ in general is the linear combination of $\rho_j\otimes \lambda_k \otimes \sigma_l$ with coefficients depending on $\bsl{r}_{\shpa}$.
In this case, we can convert the symmetry transformation of $l_\chi$ and $l_c$ indexes of $M_V(\bsl{r}_{\shpa})$ to the transformations of $\rho_j$'s and $\lambda_k$'s, respectively.
Based on \eqnref{eq:sym_d}, $\rho_j$'s and $\lambda_k$'s can be classified according to the IRs of $C_{3v}$ and parities of TR, PH and $\chi$, as shown in the second lowest and lowest parts of \tabref{tab:IR_C3v_TR_PH_chi}.
Then, the terms in $M_{V}(\bsl{r}_{\shpa})$  with certain symmetry properties can be constructed via the tensor product of the classified $\rho_j$'s, $\lambda_k$'s and $\sigma_l$'s listed in \tabref{tab:IR_C3v_TR_PH_chi}, which can further determine the number of LDOS peaks.
Similar as \secref{sec:MF_order_MFB}, the LDOS discussed here is based on the full spectrum of $M_V(\bsl{r}_{\shpa})$, of which only the half with non-negative energy is physical.
In the following, we study the LDOS at the impurity position $\bsl{r}_{\shpa}=0$ with the focus on two types of impurities: (i) non-magnetic charge impurity, and (ii) magnetic impurity with magnetization along the $(111)$ direction.


\begin{figure*}[t!]
    \centering
    \includegraphics[width=\textwidth]{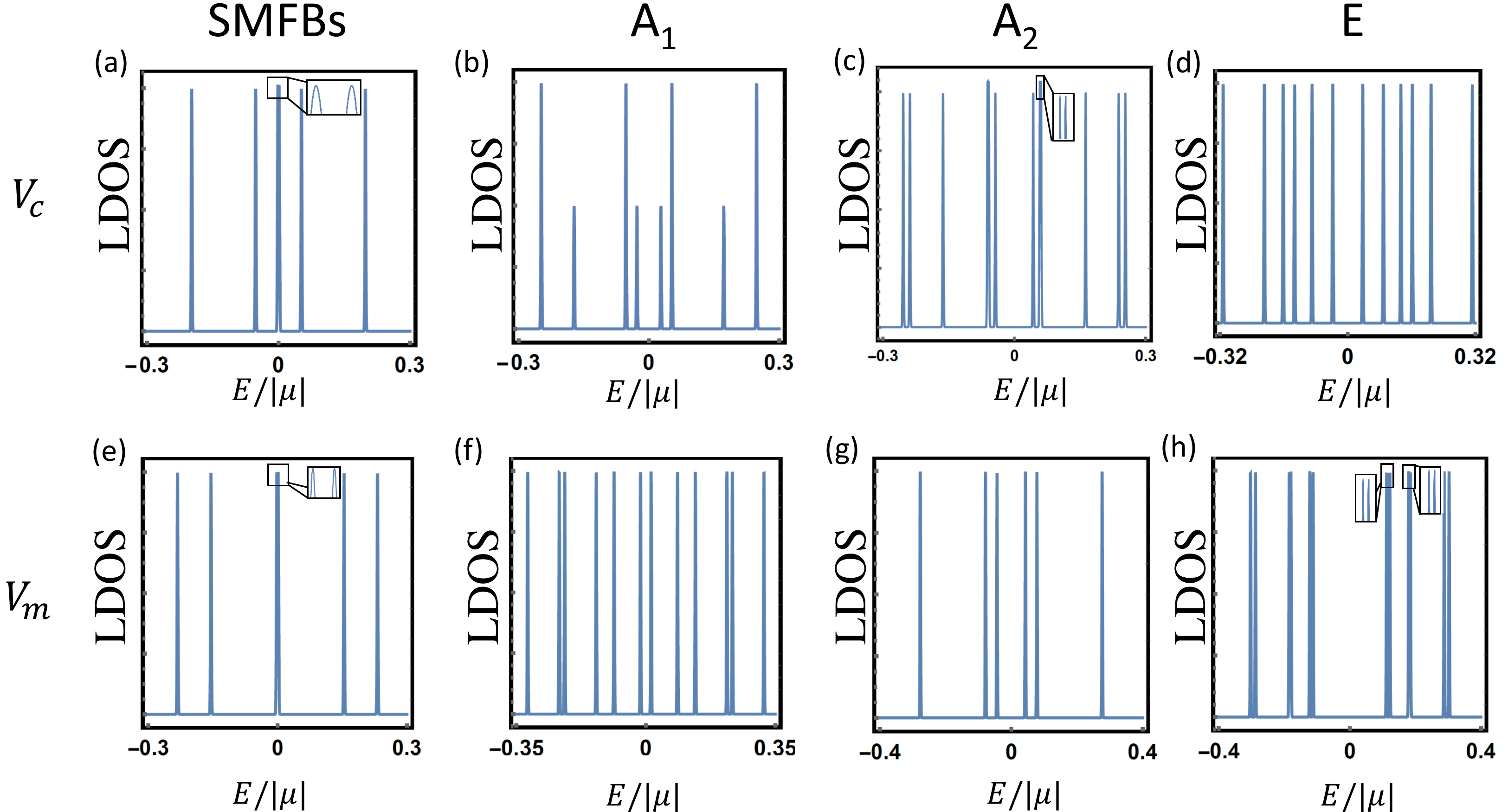}
    \caption{\label{fig:LDOS} The LDOS as a function of the energy ($E/|\mu|$) with surface impurities.
    The two rows from top to bottom are at a surface charge impurity and at a surface magnetic impurity, respectively.
The four columns from left to right correspond to no order parameters, $A_1$ order parameter, $A_2$ order parameter and $E$ order parameter, respectively.
    The broadening of each peak and the parameters choices for the orders if exist are the same as \figref{fig:LDOS_no_imp}. The potential form of the charge or magnetic impurity is shown in Appendix.\ref{app:H_d_bases}.
  The numbers on the vertical axes are again omitted.
    }
\end{figure*}

\begin{figure*}[t!]
    \centering
    \includegraphics[width=0.9\textwidth]{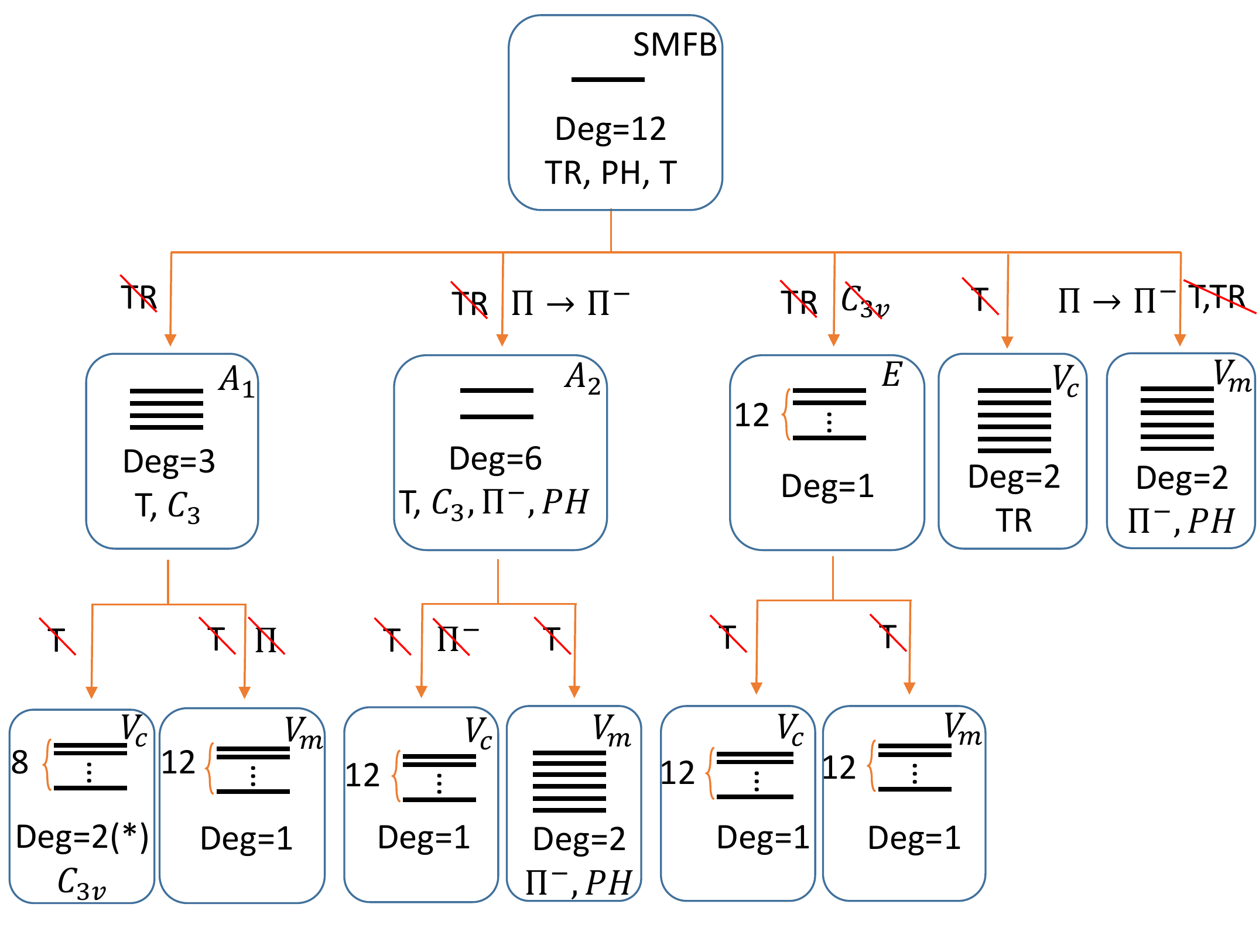}
    \caption{\label{fig:LDOS_peak_split} This graph shows how the number of LDOS peaks shown in \figref{fig:LDOS_no_imp},\ref{fig:LDOS} is determined by the symmetry.
The solid black lines indicate the LDOS peaks.
$A_1$, $A_2$ and $E$ stand for the surface order parameters, and $V_c$ and $V_m$ denote the charge and magnetic impurity, respectively.
``Deg'' indicates the symmetry protected degeneracy of the each LDOS peak, except the case marked by (*) where only half of the eight peaks have the double degeneracy.
If $Deg>1$, the line below shows the crucial symmetries that account for the degeneracy.
Here $\Pi^-$ means odd mirror parity, $T$ means the translational invariance, and the origin for the rotation $C_3$ or mirror $\Pi$ is located at the impurity center.
The red lines crossing the symmetry operations indicate the breaking of the corresponding symmetries.}
\end{figure*}

\subsection{Non-magnetic Charge Impurity}

For a charge impurity, the potential term $M_V(\bsl{r}_{\shpa}=0)=M_c$ possesses the TR symmetry $\mathcal{T}_d M^*_{c}\mathcal{T}_d^{\dagger}=M_c$, the $C_{3v}$ symmetries centered at the impurity $R_d M_c R_d^{\dagger}=M_c$ with $R\in C_{3v}$, and the chiral symmetry $\chi_d M_c \chi_d^{\dagger}=-M_c$.
(See Appendix.\ref{app:H_d_bases} for details.)
According to its symmetry properties and \tabref{tab:IR_C3v_TR_PH_chi}, the generic form of $M_c$ reads
\begin{eqnarray}
\label{eq:c_imp}
&&M_c=(\eta_1\rho_1+\eta_2\rho_2)\otimes \Lambda_1\otimes \sigma_0\\
&&+(\eta_3\rho_1+\eta_4\rho_2)\otimes \Lambda_2\otimes \sigma_0+(\eta_5\rho_1+\eta_6\rho_2)\otimes \Lambda_3\otimes \sigma_3\nonumber\\
&&+(\eta_7\rho_1+\eta_8\rho_2)\otimes (-\Lambda_{5,1}\otimes \sigma_2+\Lambda_{5,2}\otimes \sigma_1)\nonumber\ ,
\end{eqnarray}
where $\eta_{1,...,8}$ are real.
Below we examine the LDOS on a single charge impurity for SMFBs and compare the case without any order parameter to the cases with $A_1$ (\ref{eq:H_mf_A1}), $A_2$ (\ref{eq:H_mf_A2}), and $E$ (\ref{eq:H_mf_E}) order parameters.
The LDOS around the charge impurity is shown in Figs.\ref{fig:LDOS}a-d, which reveal the following features.
(1) Since PH symmetry exists in all the cases, the LDOS is always symmetric with respect to zero energy.
(2) If no order parameters exist, there are six peaks (\figref{fig:LDOS}a), given by the TR protected double degeneracy of each eigenvalue of $M_c$ according to the Kramer's degeneracy.
(3) In the presence of the $A_1$ order parameter, 8 peaks exist at the impurity (\figref{fig:LDOS}b).
The reason is the following. Since the translational invariance is absent, the modes with different $l_\chi$ or $l_c$ are coupled by the charge impurity, and the three-fold degeneracy for the pure $A_1$ order parameter case is lifted.
Moreover, the appearance of the order parameter breaks the TR symmetry, leaving only the $C_{3v}$ symmetries to protect the degeneracy.
For convenience, we choose the eigenstates of $\hat{C}_3$ rotation as the bases to make the representation $C_{3,d}$ diagonal as
\begin{equation}
\label{eq:C3t_d}
\widetilde{C}_{3,d}
=
\left(
\begin{array}{ccc}
e^{-i \frac{\pi}{3}}\mathds{1}_4 & & \\
& -\mathds{1}_4 & \\
& & e^{i \frac{\pi}{3}}\mathds{1}_4\\
\end{array}
\right)\ ,
\end{equation}
where $\mathds{1}_n$ is the $n\times n$ identity matrix.
Due to the presence of the $A_1$ order order parameter, the Hamiltonian at the charge impurity becomes $M_c+M_{A_1}$ with $M_{A_1}$ given by transforming \eqnref{eq:H_mf_A1} to the $d$ bases. (See Appendix.\ref{app:H_d_bases}.)
With the eigen-bases of $\hat{C}_3$ rotation, $M_c+M_{A_1}$ can be block diagonalized as $\text{diag}(h_1,h_2,h_3)$, where $h_1$, $h_2$ and $h_3$ are $4\times 4$ Hermitian matrices.
With the same bases, the mirror matrix $\Pi_d$ has the form
\begin{equation}
\label{eq:Pit_d}
\widetilde{\Pi}_{d}
=
\left(
\begin{array}{ccc}
 &  & U_\Pi\\
& U_\Pi & \\
U_\Pi & & \\
\end{array}
\right)
\end{equation}
with
\begin{equation}
U_\Pi=
\left(
\begin{array}{cccc}
 0 & -1 & 0 & 0 \\
 1 & 0 & 0 & 0 \\
 0 & 0 & 0 & -1 \\
 0 & 0 & 1 & 0 \\
\end{array}
\right)\ .
\end{equation}
The mirror symmetry gives $U_{\Pi} h_3 U_{\Pi}^{\dagger}=h_1$ and $U_{\Pi} h_2 U_{\Pi}^{\dagger}=h_2$, which means the eigenvalues of $h_1$ are the same as those of $h_3$.
In fact, the representations of symmetry operations show that the bases of $h_1$ and $h_3$ belong to two dimensional IRs of $C_{3v}$ while those of $h_2$ belong to one dimensional IRs of $C_{3v}$.
Therefore, $M_c+M_{A_1}$ has four doubly degenerate and four single eigenvalues, resulting in the 8 LDOS peaks.
(4) The 12 LDOS peaks exist at the impurity in the presence of the $A_2$ order parameter (\figref{fig:LDOS}c) since the translational invariance and the odd mirror parity of the $A_2$ order parameter are broken by impurity, and there are no symmetries ensuring any degeneracy.
(5) The 12 LDOS peaks at the impurity for the $E$ order parameter (\figref{fig:LDOS}d) are because no new symmetries are brought by the impurity.
Besides the above five features, the sign change of the charge does not affect the LDOS peaks since the order parameters are all chiral anti-symmetric while the charge impurity is chiral symmetric.

\subsection{Magnetic Impurity}

$M_V(\bsl{r}_{\shpa}=0)=M_m$ is still Hermitian and PH symmetric at a magnetic impurity with magnetic momentum along (111) direction.
Moreover, it is TR-odd $\mathcal{T}_d M_m^*\mathcal{T}_d^{\dagger}=-M_m$, $\hat{C}_{3}$-symmetric $C_{3,d} M_m C_{3,d}^{\dagger}=M_m$, and $\hat{\Pi}$-odd $\Pi_{d} M_m\Pi_{d}^{\dagger}=-M_m$.
(Appendix.\ref{app:H_d_bases}.)
According to the symmetry properties and \tabref{tab:IR_C3v_TR_PH_chi}, the generic form of $M_m$ reads
\begin{eqnarray}
\label{eq:m_imp_111}
&&M_m=\eta_9\rho_0\otimes \Lambda_1\otimes \sigma_3\nonumber\\
&&+\eta_{10}\rho_0\otimes \Lambda_2\otimes \sigma_3+\eta_{11}\rho_0\otimes \Lambda_3\otimes \sigma_0\nonumber\\
&&+\eta_{12}\rho_0\otimes (\Lambda_{4,2}\otimes \sigma_2+\Lambda_{4,1}\otimes \sigma_1)\nonumber\\
&&+\eta_{13}\rho_3\otimes (\Lambda_{5,2}\otimes \sigma_2+\Lambda_{5,1}\otimes \sigma_1)\nonumber\\
&&+\eta_{14}\rho_0\otimes (\Lambda_{6,2}\otimes \sigma_2+\Lambda_{6,1}\otimes \sigma_1)\ ,
\end{eqnarray}
where $\eta_{9,...,14}$ are real.
Figs.\ref{fig:LDOS}e-h show the LDOS around the magnetic impurity and reveal the following features.
(1) PH symmetry again ensures that the LDOS is always symmetric with respect to zero energy and the $E$ order parameter still has 12 LDOS peaks at the magnetic impurity since no new symmetries appear as shown in \figref{fig:LDOS}h.
(2) If no order parameters exist, there are six peaks (Figs.\ref{fig:LDOS}e), resulted from the double degeneracy given by the combination of the PH symmetry and odd $\hat{\Pi}$ parity.
It is because the combination of the PH symmetry and odd $\Pi$ parity
gives
$\Pi_{d} \mathcal{C}_d M_m \mathcal{C}_d^{\dagger} \Pi_{d}^{\dagger} = M_m$, and since $\Pi_{d} \mathcal{C}_d (\Pi_{d} \mathcal{C}_d)^*=-1$,
each eigenvalue of $M_m$ must be doubly degenerate (similar to Kramer's theorem).
(3) The original 4 peaks of the $A_1$ order are splitted into 12 peaks since the magnetic impurity breaks the translational invariance and $\hat{\Pi}$ symmetry (\figref{fig:LDOS}f).
(4) As shown in \figref{fig:LDOS}g, the 6 LDOS peaks of the magnetic impurity remain in the presence of the $A_2$ order since the PH symmetry and odd $\hat{\Pi}$ parity are not broken.
Besides the above four features, flipping the direction of the magnetic moment, i.e. $M_m\rightarrow -M_m$, does not affect the LDOS distribution in presence of the $A_1$ order parameter, since the $A_{1}$ order parameter has $\hat{\Pi}$ symmetry while $M_m$ has odd $\hat{\Pi}$ parity.

\subsection{Summary for Impurity Effect}

To sum up, the number of LDOS peaks at a charge impurity or a magnetic impurity with magnetic moment in $(111)$ direction is 6 or 6 for no order parameters, 8 or 12 for the $A_1$ order parameter, 12 or 6 for the $A_2$ order parameter, and 12 or 12 for the $E$ order parameter, respectively, as summarized in \figref{fig:LDOS_peak_split}.
Combining the above results with the LDOS peaks without impurity given in \secref{sec:MF_order_MFB}, it is more than enough to identify the order parameters in our system.
In the above analysis, we adopt the approximation (\ref{eq:appro_local_d}), only consider translationally invariant order parameters that are $\bsl{k}_\shpa$-independent in each surface mode region, and assume the surface mode wavefunctions are $\bsl{k}_\shpa$-independent in each surface mode region to deal with the impurity.
Those approximations neglect high-order effects which typically can only broaden the LDOS peaks without affecting the qualitative result.

\section{Conclusion and Discussion}
\label{sec:conclusion}
In this work, we studied the energy spectrum (or LDOS) of the SMFBs localized on (111) surface of the half-Heusler SCs with translationally invariant order parameters or magnetic/non-magnetic impurities based on the Luttinger model with singlet-quintet mixing.
Our work demonstrates that the zero-bias peak of SMFBs can be split to reveal a rich peak structure when different types of order parameters induced by interaction or magnetic/non-magnetic impurities are introduced.
Such peak structure can be viewed as a fingerprint to distinguish different types of order parameters in the standard STM experiments.
In addition, we notice that the SMFBs induced by singlet-septet mixing proposed in \refcite{Brydon2016j=3/2SC} possess six patches without any additional pseudospin degeneracy in the surface Brillouin zone (see Fig.\,5a and the discussion in \refcite{Timm2017nodalj=3/2SC}).
Due to the different number of degeneracy, we expect the peak structures given by the order parameters and magnetic/non-magnetic impurities will be different in two cases, which thereby may help distinguish the singlet-quintet mixing from the singlet-septet mixing in experiments.

\section{Acknowledgement}
We acknowledge the helpful discussion with C.Wu.
J.Y thanks Yang Ge, Rui-Xing Zhang, Jian-Xiao Zhang and Tongzhou Zhao for helpful discussion.
We acknowledge the support of the Office of Naval Research (Grant No. N00014-18-1-2793), Kaufman New Initiative research grant KA2018-98553 of the Pittsburgh Foundation and the U.S. Department of Energy (Grant No.~DESC0019064).

\begin{appendices}
\section{Convention and Expressions}
\label{app:conv_expn}
The Fourier transformation of creation operators in the continuous limit reads
\begin{equation}
c^{\dagger}_{\bsl{r}}=\frac{1}{\sqrt{\mathcal{V}}}\sum_{\bsl{k}}e^{-i\bsl{k}\cdot\bsl{r}}c^{\dagger}_{\bsl{k}}\ ,
\end{equation}
where $\mathcal{V}$ is the total volume of the entire space.

The five d-orbital cubic harmonics read \cite{Murakami2004SU2}
\begin{equation}
\left\{
\begin{array}{l}
g_{\bsl{k},1}=\sqrt{3} k_y k_z\\
g_{\bsl{k},2}=\sqrt{3} k_z k_x\\
g_{\bsl{k},3}=\sqrt{3} k_x k_y\\
g_{\bsl{k},4}=\frac{\sqrt{3}}{2} (k_x^2-k_y^2)\\
g_{\bsl{k},5}=\frac{1}{2}(2 k_z^2-k_x^2-k_y^2)\\
\end{array}
\right..
\end{equation}

The $j=\frac{3}{2}$ angular momentum matrices are \cite{Winkler2003SOC}
\begin{equation}
J_x=\left(
\begin{array}{cccc}
 0 & \frac{\sqrt{3}}{2} & 0 & 0 \\
 \frac{\sqrt{3}}{2} & 0 & 1 & 0 \\
 0 & 1 & 0 & \frac{\sqrt{3}}{2} \\
 0 & 0 & \frac{\sqrt{3}}{2} & 0 \\
\end{array}
\right)
\end{equation}
\begin{equation}
J_y=\left(
\begin{array}{cccc}
 0 & -\frac{i \sqrt{3}}{2} & 0 & 0 \\
 \frac{i \sqrt{3}}{2} & 0 & -i & 0 \\
 0 & i & 0 & -\frac{i \sqrt{3}}{2}  \\
 0 & 0 & \frac{i \sqrt{3}}{2} & 0 \\
\end{array}
\right)
\end{equation}
\begin{equation}
J_z=\left(
\begin{array}{cccc}
 \frac{3}{2} & 0 & 0 & 0 \\
 0 & \frac{1}{2} & 0 & 0 \\
 0 & 0 & -\frac{1}{2} & 0 \\
 0 & 0 & 0 & -\frac{3}{2} \\
\end{array}
\right).
\end{equation}

The five Gamma matrices are \cite{Murakami2004SU2}
\begin{equation}
\left\{
\begin{array}{l}
\Gamma^1=\frac{1}{\sqrt{3}} (J_y J_z+J_z J_y)\\
\Gamma^2=\frac{1}{\sqrt{3}} (J_z J_x+J_x J_z)\\
\Gamma^3=\frac{1}{\sqrt{3}} (J_x J_y+J_y J_x)\\
\Gamma^4=\frac{1}{\sqrt{3}} (J_x^2-J_y^2)\\
\Gamma^5=\frac{1}{3} (2 J_z^2-J_x^2-J_y^2)\\
\end{array}
\right..
\end{equation}
Clearly, $\{\Gamma^a,\Gamma^b\}=2\delta_{ab}\Gamma^0$ where $\Gamma^0$ is the 4 by 4 identity matrix.

\begin{table}[H]
  \centering
  \begin{tabular}{|c|c|c|c|}
\hline
  $C_{3v}$ & $\mathds{1}$ & $C_3$ & $\Pi$ \\ \hline
    $A_1$ & 1 & 1 & 1 \\ \hline
   $A_2$ & 1 & 1 & -1 \\ \hline
   $E$ & 2 & -1 & 0 \\ \hline
    \end{tabular}
    \caption{\label{tab:cha_C3v} Character table of $C_{3v}$. Here $\mathds{1}$ means identity operation.\cite{Aroyo2006BilbaoIR}}
    \end{table}

\begin{table}[H]
  \centering
  \begin{tabular}{|c|m{6 cm}|c|}
\hline
  $C_{3v}$ & & TR \\ \hline
    $A_1$ & $n_0=\Gamma_0 $ & $+$ \\ \hline
    $A_1$ & $n_1=\frac{1}{\sqrt{3}}(\Gamma_1+\Gamma_2+\Gamma_3) $ & $+$ \\ \hline
    $A_1$ & $n_2=\frac{1}{\sqrt{3}}(V_x+V_y+V_z) $ & $-$ \\ \hline
    $A_2$ & $n_3= J_{xyz} $ & $-$ \\ \hline
    $A_2$ & $n_4=\frac{1}{\sqrt{3}}(J_x+J_y+J_z)$ & $-$ \\ \hline
    $A_2$ & $n_5=\frac{1}{\sqrt{3}}(P_x+P_y+P_z)$ & $-$ \\ \hline
    $E$ &  $\bsl{n}_{6}=$ $( \frac{1}{\sqrt{6}}( \Gamma_1+\Gamma_2-2\Gamma_3 ), \frac{1}{\sqrt{2}}( -\Gamma_1+\Gamma_2 ))$  & $+$ \\ \hline
    $E$ &  $\bsl{n}_{7}=$ $(\Gamma_5, \Gamma_4)$  & $+$ \\ \hline
    $E$ &  $\bsl{n}_{8}=$ $(\frac{1}{\sqrt{2}}(J_x - J_y ), \frac{1}{\sqrt{6}}(J_x+J_y -2 J_z ) )$  & $-$ \\ \hline
    $E$ &  $\bsl{n}_{9}=$ $(\frac{1}{\sqrt{2}}( P_x-P_y ), \frac{1}{\sqrt{6}}(P_x+P_y -2 P_z) )$  & $-$ \\ \hline
    $E$ &   $\bsl{n}_{10}=$ $( \frac{1}{\sqrt{6}}( V_x+V_y-2V_z ), \frac{1}{\sqrt{2}}( -V_x+V_y ) )$  & $-$ \\ \hline
    \end{tabular}
    \caption{\label{tab:n_class}Expressions of $n_i$ in \eqnref{eq:Mt_Dt}. $P_i=J_i^3-41 J_i/20$, $V_x=\frac{1}{2}\{J_x,J_y^2-J_z^2\}$,
$V_y=\frac{1}{2}\{J_y,J_z^2-J_x^2\}$,
$V_z=\frac{1}{2}\{J_z,J_x^2-J_y^2\}$ and $J_{xyz}=J_x J_y J_z+J_z J_y J_x$.}
    \end{table}

\begin{table}[H]
  \centering
  \begin{tabular}{|c|m{6 cm}|c|}
\hline
  $C_{3v}$ & & TR \\ \hline
    $A_1$ & $N_{1}(\bsl{k}_{\shpa})=\delta^{\chi}_{\bsl{k}_{\shpa}}\sigma_0 $ & $-$ \\ \hline
    $A_1$ & $N_{2}(\bsl{k}_{\shpa})=\delta^{E_1,+}_{\bsl{k}_{\shpa}}(-\sigma_2) +\delta^{E_2,+}_{\bsl{k}_{\shpa}}\sigma_1 $ & $-$ \\ \hline
    $A_2$ & $N_{3}(\bsl{k}_{\shpa})=\sigma_3 $ & $-$ \\ \hline
    $A_2$ & $N_{4}(\bsl{k}_{\shpa})=-\delta^{E_2,+}_{\bsl{k}_{\shpa}}(-\sigma_2) +\delta^{E_1,+}_{\bsl{k}_{\shpa}}\sigma_1 $ & $-$ \\ \hline
    $E$ &  $\bsl{N}_{5}(\bsl{k}_{\shpa})=$ $(\delta^{E_1,-}_{\bsl{k}_{\shpa}}\sigma_0 , \delta^{E_2,-}_{\bsl{k}_{\shpa}}\sigma_0 )$  & $-$ \\ \hline
    $E$ &  $\bsl{N}_{6}(\bsl{k}_{\shpa})=$ $(-\delta^{E_2,+}_{\bsl{k}_{\shpa}}\sigma_3 , \delta^{E_1,+}_{\bsl{k}_{\shpa}}\sigma_3 )$  & $-$ \\ \hline
    $E$ &  $\bsl{N}_{7}(\bsl{k}_{\shpa})=$ $(-\sigma_2 ,\sigma_1 )$  & $-$ \\ \hline
    $E$ & $\bsl{N}_{8}(\bsl{k}_{\shpa})=$ $(-\delta^{E_1,+}_{\bsl{k}_{\shpa}}(-\sigma_2) +\delta^{E_2,+}_{\bsl{k}_{\shpa}}\sigma_1 $, $\delta^{E_1,+}_{\bsl{k}_{\shpa}}\sigma_1  + \delta^{E_2,+}_{\bsl{k}_{\shpa}}(-\sigma_2) )$ & $-$ \\ \hline
    \end{tabular}
    \caption{\label{tab:N}Expressions of $N_i$ in \eqnref{eq:H_mf_A1}, \eqnref{eq:H_mf_A2} and \eqnref{eq:H_mf_E}.}
    \end{table}

The list of Gell-Mann matrices\cite{GellMannMatrices1962}
\begin{equation}
\begin{array}{cc}
\lambda_1 =
\left(
\begin{array}{ccc}
 0 & 1 & 0 \\
 1 & 0 & 0 \\
 0 & 0 & 0 \\
\end{array}
\right) &
\lambda_2=\left(
\begin{array}{ccc}
 0 & -i & 0 \\
 i & 0 & 0 \\
 0 & 0 & 0 \\
\end{array}
\right) \\
\lambda_3=\left(
\begin{array}{ccc}
 1 & 0 & 0 \\
 0 & -1 & 0 \\
 0 & 0 & 0 \\
\end{array}
\right)&
\lambda_4=\left(
\begin{array}{ccc}
 0 & 0 & 1 \\
 0 & 0 & 0 \\
 1 & 0 & 0 \\
\end{array}
\right)
\\
\lambda_5=\left(
\begin{array}{ccc}
 0 & 0 & -i \\
 0 & 0 & 0 \\
 i & 0 & 0 \\
\end{array}
\right)
&
\lambda_6=\left(
\begin{array}{ccc}
 0 & 0 & 0 \\
 0 & 0 & 1 \\
 0 & 1 & 0 \\
\end{array}
\right)\\
\lambda_7=\left(
\begin{array}{ccc}
 0 & 0 & 0 \\
 0 & 0 & -i \\
 0 & i & 0 \\
\end{array}
\right)&
\lambda_8=\frac{1}{\sqrt{3}}\left(
\begin{array}{ccc}
 1 & 0 & 0 \\
 0 & 1 & 0 \\
 0 & 0 & -2 \\
\end{array}
\right)
\end{array}\ .
\end{equation}
And $\lambda_0$ is defined as the $3\times 3$ identity matrix.

\section{Representations of Symmetry Operators}
\label{app:rep_sym}
In this section, we show the representation of symmetry operators on the $c^{\dagger}_{\bsl{k}}$ bases and the Nambu bases.
Before showing the representation, we define the following notations:
$\hat{P}_F$ is the fermion parity operator,
$\hat{T}_{\bsl{x}}$ with $\bsl{x}\in \mathds{R}^3$ is a generic translation operator,
the generators of $O_h$ group $\hat{C}_{3}$, $\hat{P}$, $\hat{C}_{4}$ and $\hat{\Pi}$ are 3-fold rotations along $(111)$, inversion, 4-fold rotation along $(001)$ and mirror perpendicular to $(1\bar{1}0)$, respectively, and
$\hat{\mathcal{T}}$ is the time-reversal operator.
Representations of $O(3)$ are not shown here since we only care about the $c_1\neq c_2$ case.

\subsubsection{The $c^{\dagger}_{\mathbf{k}}$ Bases}
\begin{equation}
\hat{P}_F c^{\dagger}_{\bsl{k}} \hat{P}_F^{-1}=-c^{\dagger}_{\bsl{k}}\ ,\ \hat{P}_F c_{\bsl{k}} \hat{P}_F^{-1}=-c_{\bsl{k}}\ ,
\end{equation}
\begin{equation}
\hat{T}_{\bsl{x}} c^{\dagger}_{\bsl{k}} \hat{T}_{\bsl{x}}^{-1}=e^{-i \bsl{k}\cdot\bsl{x}}c^{\dagger}_{\bsl{k}}\ ,\
\hat{T}_{\bsl{x}} c_{\bsl{k}} \hat{T}_{\bsl{x}}^{-1}=e^{i \bsl{k}\cdot\bsl{x}}c_{\bsl{k}}\ ,
\end{equation}
\begin{equation}
\hat{C}_{3} c^{\dagger}_{\bsl{k}} \hat{C}_{3}^{-1}=c^{\dagger}_{C_{3}\bsl{k}}C_3\ ,\
\hat{C}_{3} c_{\bsl{k}} \hat{C}_{3}^{-1}=C_3^{\dagger}c_{C_{3}\bsl{k}}\ ,
\end{equation}
\begin{equation}
\hat{P} c^{\dagger}_{\bsl{k}} \hat{P}^{-1}=-c^{\dagger}_{-\bsl{k}}\ ,\
\hat{P} c_{\bsl{k}} \hat{P}^{-1}=-c_{-\bsl{k}}\ ,\
\end{equation}
\begin{equation}
\hat{C}_{4} c^{\dagger}_{\bsl{k}} \hat{C}_{4}^{-1}=c^{\dagger}_{C_{4}\bsl{k}}C_4\ ,\
\hat{C}_{4} c_{\bsl{k}} \hat{C}_{4}^{-1}=C_4^{\dagger} c_{C_{4}\bsl{k}}\ ,
\end{equation}
\begin{equation}
\hat{\Pi} c^{\dagger}_{\bsl{k}} \hat{\Pi}^{-1}=c^{\dagger}_{\Pi\bsl{k}}\Pi\ ,\
\hat{\Pi} c_{\bsl{k}} \hat{\Pi}^{-1}=\Pi^{\dagger} c_{\Pi \bsl{k}}\ ,
\end{equation}
\begin{equation}
\hat{\mathcal{T}} c^{\dagger}_{\bsl{k}} \hat{\mathcal{T}}^{-1}=c^{\dagger}_{-\bsl{k}}\gamma\ ,\
\hat{\mathcal{T}} c_{\bsl{k}} \hat{\mathcal{T}}^{-1}=\gamma^{\dagger} c_{-\bsl{k}}\ ,
\end{equation}
where
$C_3= \exp(-i\frac{J_x+J_y+J_z}{\sqrt{3}}\frac{2\pi}{3})$, $C_3\bsl{k}=(k_z,k_x,k_y)$,
$C_4= \exp(-i J_z \frac{2\pi}{4})$,
$C_4\bsl{k}=(-k_y,k_x,k_z)$,
$\Pi=-\exp(-i\frac{J_x-J_y}{\sqrt{2}}\frac{2\pi}{2})$ and
$\Pi \bsl{k}=(k_y,k_x,k_z)$.

\subsubsection{The Nambu Bases}
\begin{equation}
\hat{P}_F \Psi^{\dagger}_{\bsl{k}} \hat{P}_F^{-1}=-\Psi^{\dagger}_{\bsl{k}}\ ,\ \hat{P}_F \Psi_{\bsl{k}} \hat{P}_F^{-1}=-\Psi_{\bsl{k}}\ ,
\end{equation}
\begin{equation}
\hat{T}_{\bsl{x}} \Psi^{\dagger}_{\bsl{k}} \hat{T}_{\bsl{x}}^{-1}=e^{-i \bsl{k}\cdot\bsl{x}}\Psi^{\dagger}_{\bsl{k}}\ ,\
\hat{T}_{\bsl{x}} \Psi_{\bsl{k}} \hat{T}_{\bsl{x}}^{-1}=e^{i \bsl{k}\cdot\bsl{x}}\Psi_{\bsl{k}}\ ,
\end{equation}
\begin{equation}
\hat{C}_{3} \Psi^{\dagger}_{\bsl{k}} \hat{C}_{3}^{-1}=\Psi^{\dagger}_{C_{3}\bsl{k}}\widetilde{C}_3\ ,\
\hat{C}_{3} \Psi_{\bsl{k}} \hat{C}_{3}^{-1}=\widetilde{C}_3^{\dagger}\Psi_{C_{3}\bsl{k}}\ ,
\end{equation}
\begin{equation}
\hat{P} \Psi^{\dagger}_{\bsl{k}} \hat{P}^{-1}=-\Psi^{\dagger}_{-\bsl{k}}\ ,\
\hat{P} \Psi_{\bsl{k}} \hat{P}^{-1}=-\Psi_{-\bsl{k}}\ ,\
\end{equation}
\begin{equation}
\hat{C}_{4} \Psi^{\dagger}_{\bsl{k}} \hat{C}_{4}^{-1}=\Psi^{\dagger}_{C_{4}\bsl{k}}\widetilde{C}_4\ ,\
\hat{C}_{4} \Psi_{\bsl{k}} \hat{C}_{4}^{-1}=\widetilde{C}_4^{\dagger} \Psi_{C_{4}\bsl{k}}\ ,
\end{equation}
\begin{equation}
\hat{\Pi} \Psi^{\dagger}_{\bsl{k}} \hat{\Pi}^{-1}=\Psi^{\dagger}_{\Pi\bsl{k}}\widetilde{\Pi}\ ,\
\hat{\Pi} \Psi_{\bsl{k}} \hat{\Pi}^{-1}=\widetilde{\Pi}^{\dagger} \Psi_{\Pi\bsl{k}}\ ,
\end{equation}
\begin{equation}
\hat{\mathcal{T}} \Psi^{\dagger}_{\bsl{k}} \hat{\mathcal{T}}^{-1}=\Psi^{\dagger}_{-\bsl{k}}\mathcal{T}\ ,\
\hat{\mathcal{T}} \Psi_{\bsl{k}} \hat{\mathcal{T}}^{-1}=\mathcal{T}^{\dagger} \Psi_{-\bsl{k}}\ ,
\end{equation}
where
$\widetilde{C}_3=\text{diag}(C_3,C_3^*)$,
$\widetilde{C}_4= \text{diag}(C_4,C_4^*)$,
$\widetilde{\Pi}= \text{diag}(\Pi,\Pi^*)$ and
$\mathcal{T}=\text{diag}(\gamma, \gamma^*)$.
$\widetilde{C}_3$, $\widetilde{\Pi}$, $\mathcal{T}K$ and $\mathcal{C}K$ commute with each other, where $K$ is the complex conjugate operation.
$\chi$ anti-commutes with $\mathcal{T}K$ and $\mathcal{C}K$ and commutes with $\widetilde{C}_3$ and $\widetilde{\Pi}$.

\section{Surface Majorana Flat Bands}
\label{app:surf_modes}

\subsubsection{Existence of Surface Zero Modes}
Due to the topological invariant $N_{w}=\pm 2$ at each non-trivial $\bsl{k}_{\shpa}$, we expect two boundary modes at each non-trivial $\bsl{k}_{\shpa}$ on one surface of our model. \cite{yu2017Singlet-Quintetj=3/2SC}
Therefore, we consider a semi-infinite version of \eqnref{eq:H_BdG} ($x_{\perp}<0$) with open boundary condition at $x_{\perp}=0$, where $x_{\perp}$ is the position on $(111)$ axis. The corresponding Hamiltonian reads
\begin{eqnarray}
\label{eq:H_BdG_perp}
&&H_{\perp}=\frac{1}{2}\sum_{\bsl{k}_{\shpa}}\int_{-\infty}^0 dx_{\perp}\Psi_{\bsl{k}_{\shpa}, x_{\perp}}^{\dagger}h_{BdG}(\bsl{k}_{\shpa},-i\partial_{x_{\perp}})\Psi_{\bsl{k}_{\shpa},x_{\perp}}\nonumber\\
&& +\sum_{\bsl{k}_{\shpa}}\int^{+\infty}_0 dx_{\perp} E_{\infty} c^{\dagger}_{\bsl{k}_{\shpa}, x_{\perp}} c_{\bsl{k}_{\shpa}, x_{\perp}} +const.\ ,
\end{eqnarray}
where $c^{\dagger}_{\bsl{k}_{\shpa}, x_{\perp}}=\frac{1}{\sqrt{L_{\perp}}}\sum_{k_{\perp}}e^{-i k_{\perp} x_{\perp}}c^{\dagger}_{\bsl{k}}$ with $L_{\perp}$ the length along the $(111)$ direction of the entire space, $h_{BdG}(\bsl{k}_{\shpa},-i\partial_{x_{\perp}})$ is obtained by replacing $k_{\perp}$ in  $h_{BdG}(\bsl{k})$ by $-i\partial_{x_{\perp}}$, $\Psi^{\dagger}_{\bsl{k}_{\shpa}, x_{\perp}}=(c_{\bsl{k}_{\shpa}, x_{\perp}}^{\dagger},c_{-\bsl{k}_{\shpa}, x_{\perp}}^{T})$, and $E_{\infty}\rightarrow +\infty$ is for the open boundary condition.
For such a semi-infinite system, the translation symmetry in the $(111)$ direction, the inversion symmetry and the 4-fold rotational symmetry along $(001)$ are broken.
The Hamiltonian $h_{BdG}(\bsl{k}_{\shpa},-i\partial_{x_{\perp}})$ still has PH, TR, chiral and $C_{3v}$ symmetries 
$-\mathcal{C}[h_{BdG}(-\bsl{k}_{\shpa},-i\partial_{x_{\perp}})]^*\mathcal{C}^{\dagger}=h_{BdG}(\bsl{k}_{\shpa},-i\partial_{x_{\perp}})$, $\mathcal{T}[h_{BdG}(-\bsl{k}_{\shpa},-i\partial_{x_{\perp}})]^*\mathcal{T}^{\dagger}=h_{BdG}(\bsl{k}_{\shpa},-i\partial_{x_{\perp}})$, $-\chi h_{BdG}(\bsl{k}_{\shpa},-i\partial_{x_{\perp}})\chi^{\dagger}=h_{BdG}(\bsl{k}_{\shpa},-i\partial_{x_{\perp}})$ and $\widetilde{R} h_{BdG}(R^{-1}\bsl{k}_{\shpa},-i\partial_{x_{\perp}})\widetilde{R}^{\dagger}=h_{BdG}(\bsl{k}_{\shpa},-i\partial_{x_{\perp}})$, respectively, where $R=C_3, \Pi$. In addition, the PH symmetry requires $\mathcal{C}(\Psi^{\dagger}_{-\bsl{k}_{\shpa}, x_{\perp}})^T=\Psi_{\bsl{k}_{\shpa}, x_{\perp}}$ and the commutation relation is
\begin{eqnarray}
&&\{\Psi^{\dagger}_{\bsl{k}_{\shpa}, x_{\perp},\alpha,s},\Psi_{\bsl{k}_{\shpa}', x_{\perp}',\alpha',s'} \}=\delta_{\bsl{k}_{\shpa},\bsl{k}_{\shpa}'}\delta(x_{\perp}-x_{\perp}')\delta_{\alpha\alpha'}\delta_{s s'}\\
&&\{\Psi^{\dagger}_{\bsl{k}_{\shpa}, x_{\perp},\alpha,s},\Psi^{\dagger}_{\bsl{k}_{\shpa}', x_{\perp}',\alpha',s'} \}=\delta_{\bsl{k}_{\shpa},-\bsl{k}_{\shpa}'}\delta(x_{\perp}-x_{\perp}')(\tau_x)_{\alpha\alpha'}\delta_{s s'}\nonumber\ ,
\end{eqnarray}
where $\alpha,\alpha'=1,2$ stand for the particle-hole index and $s,s'$ are spin index of the $j=3/2$ fermion.

The surface mode with zero energy $b^{\dagger}_{\bsl{k}_{\shpa}}$ of $H_{\perp}$ in \eqnref{eq:H_BdG_perp} is defined as
\begin{equation}
b^{\dagger}_{\bsl{k}_{\shpa}}=\int_{-\infty}^0 d x_{\perp} \Psi_{\bsl{k}_{\shpa},x_{\perp}}^{\dagger} v_{\bsl{k}_{\shpa},x_{\perp}},
\end{equation}
 which satisfies $[H_{\perp},b^{\dagger}_{\bsl{k}_{\shpa}}]=0$ and $v_{\bsl{k}_{\shpa},0}=v_{\bsl{k}_{\shpa},-\infty}=0$.
With the PH symmetry and the commutation relation, the equation $[H_{\perp},b^{\dagger}_{\bsl{k}_{\shpa}}]=0$ can be simplified as
\begin{equation}
\label{eq:surf_modes_ori}
h_{BdG}(\bsl{k}_{\shpa},-i\partial_{x_{\perp}}) v_{\bsl{k}_{\shpa},x_{\perp}}=0\ .
\end{equation}
Now we try to figure out the properties of the solution.
First, transform the above equation to chiral eigen-bases:
\begin{equation}
\label{eq:surf_modes}
U^{\dagger}_{\chi} h_{BdG}(\bsl{k}_{\shpa},-i\partial_{x_{\perp}}) U_{\chi} U^{\dagger}_{\chi}v_{\bsl{k}_{\shpa},x_{\perp}}=0\ ,
\end{equation}
where
\begin{equation}
U_{\chi}
=
\frac{1}{\sqrt{2}}
\left(
\begin{matrix}
\mathds{1}_4 & \mathds{1}_4\\
i\gamma & -i\gamma\\
\end{matrix}
\right)
\end{equation}
is the unitary matrix that diagonalizes $\chi$:
\begin{equation}
U^{\dagger}_{\chi} \chi U_{\chi}=
\left(
\begin{matrix}
\mathds{1}_4 & \\
 & -\mathds{1}_4\\
\end{matrix}
\right)
\end{equation}
,
\begin{equation}
U^{\dagger}_{\chi} h_{BdG}(\bsl{k}_{\shpa},-i\partial_{x_{\perp}}) U_{\chi}=
\left(
\begin{matrix}
 & q(\bsl{k}_{\shpa},-i\partial_{x_{\perp}})\\
[q(\bsl{k}_{\shpa},i\partial_{x_{\perp}})]^{\dagger} & \\
\end{matrix}
\right)\ ,
\end{equation}
and
\begin{equation}
q(\bsl{k}_{\shpa},-i\partial_{x_{\perp}})=h(\bsl{k}_{\shpa},-i\partial_{x_{\perp}})-i \Delta(\bsl{k}_{\shpa},-i\partial_{x_{\perp}})\gamma\ .
\end{equation}
The TR and PH matrices in the chiral representation read
\begin{equation}
U^{\dagger}_{\chi}\mathcal{T}U^{*}_{\chi}=\left(
\begin{matrix}
 & \gamma\\
\gamma & \\
\end{matrix}
\right)
\end{equation}
and
\begin{equation}
U^{\dagger}_{\chi}\mathcal{C}U^{*}_{\chi}=\left(
\begin{matrix}
 & i \gamma\\
-i \gamma & \\
\end{matrix}
\right)\ .
\end{equation}
In the chiral representation, both TR and PH symmetries give the same condition on $q$:
\begin{equation}
\gamma [q(-\bsl{k}_{\shpa},i\partial_{x_{\perp}})]^{T} \gamma^{\dagger}=q(\bsl{k}_{\shpa},-i\partial_{x_{\perp}})\ .
\end{equation}
By defining $U^{\dagger}_{\chi}v_{\bsl{k}_{\shpa},x_{\perp}}=(u^T_{\bsl{k}_{\shpa},x_{\perp}},w^T_{\bsl{k}_{\shpa},x_{\perp}})^T$ with $u$($w$) corresponding to chiral eigen-wavefunction with chiral eigenvalues $1$($-1$), \eqnref{eq:surf_modes} can be expressed as
\begin{equation}
\label{eq:surf_modes_q}
\left\{
\begin{array}{l}
q(\bsl{k}_{\shpa},-i\partial_{x_{\perp}})w_{\bsl{k}_{\shpa},x_{\perp}}=0\\
q^{\dagger}(\bsl{k}_{\shpa},-i\partial_{x_{\perp}})u_{\bsl{k}_{\shpa},x_{\perp}}=0
\end{array}
\right. \ .
\end{equation}
Since $h_{BdG}(-\bsl{k}_{\shpa},i\partial_{x_{\perp}})=h_{BdG}(\bsl{k}_{\shpa},-i\partial_{x_{\perp}})$ originated from the bulk inversion symmetry, we have $q(\bsl{k}_{\shpa},-i\partial_{x_{\perp}})=q(-\bsl{k}_{\shpa},i\partial_{x_{\perp}})$.
Combined with TR, the equation of $u$ in \eqnref{eq:surf_modes_q} can be transformed to
\begin{equation}
q(\bsl{k}_{\shpa},-i\partial_{x_{\perp}})\gamma^T u^*_{\bsl{k}_{\shpa},-x_{\perp}}=0\ .
\end{equation}
Since $u_{\bsl{k}_{\shpa},x_{\perp}}=0$ for $x_{\perp}=0,-\infty$ which means $\gamma^T u^*_{\bsl{k}_{\shpa},-x_{\perp}}=0$ for $x_{\perp}=0,+\infty$, the above equation is the same as the equation of $w$ except that the open boundary conditions are at $x_{\perp}=0,+\infty$.
Therefore, we can solve the equation of $w$ in \eqnref{eq:surf_modes_q}, i.e.
\begin{equation}
\label{eq:surf_modes_q_w}
q(\bsl{k}_{\shpa},-i\partial_{x_{\perp}})w_{\bsl{k}_{\shpa},x_{\perp}}=0 \ ,
\end{equation}
with $w_{\bsl{k}_{\shpa},0}=w_{\bsl{k}_{\shpa},-\infty}=0$ to have the solutions of $w$ and with $w_{\bsl{k}_{\shpa},0}=w_{\bsl{k}_{\shpa},\infty}=0$ to have the solutions of $u$ by $u_{\bsl{k}_{\shpa},x_{\perp}}=\gamma w^*_{\bsl{k}_{\shpa},-x_{\perp}}$.

With the ansatz $w_{\bsl{k}_{\shpa},x_{\perp}}=e^{\lambda x_{\perp}}\bar{w}_{\bsl{k}_{\shpa}}$, the \eqnref{eq:surf_modes_q_w} becomes
\begin{equation}
\label{eq:surf_modes_q_w_ansatz}
q(\bsl{k}_{\shpa},-i\lambda)\bar{w}_{\bsl{k}_{\shpa}}=0\
\end{equation}
with the solution determined by the octic equation $\text{det}[q(\bsl{k}_{\shpa},-i\lambda)]=0$ for $\lambda$.
The equation has 4 double roots $\lambda_{1,2,3,4}$ since $\text{det}[q(\bsl{k}_{\shpa},-i\lambda)]$ can be written in the form of the square of certain function, $\text{det}[q(\bsl{k}_{\shpa},-i\lambda)]=[\widetilde{q}(\bsl{k}_{\shpa},-i\lambda)]^2$.\cite{yu2017Singlet-Quintetj=3/2SC}
In addition, since $\widetilde{q}(\bsl{k}_{\shpa},-i\lambda)$ does not have $\lambda^3$ term, the sum of $\lambda_{1,2,3,4}$ is zero.
Each double root $\lambda_i$ can give two orthogonal solutions $\bar{w}_{\bsl{k}_{\shpa},i,j}$ of \eqnref{eq:surf_modes_q_w_ansatz} with $i=1,2,3,4$ and $j=1,2$. Then the general solution of \eqnref{eq:surf_modes_q_w} without boundary condition reads
\begin{equation}
w_{\bsl{k}_{\shpa},x_{\perp}}=\sum_{i=1}^4\sum_{j=1}^2 b_{ij} e^{\lambda_i x_{\perp}}\bar{w}_{\bsl{k}_{\shpa},i,j}\ .
\end{equation}
Now let us impose the boundary condition.
$w_{\bsl{k}_{\shpa},\infty}=0$ or $w_{\bsl{k}_{\shpa},-\infty}=0$ requires $Re[\lambda_i]<0$ or $Re[\lambda_i]>0$, respectively, and $w_{\bsl{k}_{\shpa},0}=0$ requires $\sum_{i,j} b_{ij} \bar{w}_{\bsl{k}_{\shpa},i,j}=0$.
Since the sum of the four $\lambda_i$'s is zero, it is impossible to have four $Re[\lambda_i]$'s with the same sign.
If only two $Re[\lambda_i]$'s have the same sign, there will be typically no solutions, since the corresponding four four-component $\bar{w}_{\bsl{k}_{\shpa},i,j}$'s typically can not be linearly dependent.
If three $\lambda_i$'s satisfy $Re[\lambda_i]>0$($Re[\lambda_i]<0$), there are six corresponding four-component $\bar{w}_{\bsl{k}_{\shpa},i,j}$'s, resulting in two solutions to $w$($u$) corresponding to two surface zero modes $v_{\bsl{k}_{\shpa},x_{\perp}}=U_{\chi}(0,w^T_{\bsl{k}_{\shpa},x_{\perp}})^T$($v_{\bsl{k}_{\shpa},x_{\perp}}=U_{\chi}(u^T_{\bsl{k}_{\shpa},x_{\perp}},0$)) with chiral eigenvalue $-1$($1$).
Therefore, the generic number of surface zero modes at a fixed $\bsl{k}_{\shpa}$ on one surface, if exist, is two and those two modes are chiral eigenstates of the same chiral eigenvalues.
\subsubsection{Symmetries of Surface Zero Modes}
Now we will show the symmetry properties of the surface zero modes.
We take $v_{i,\bsl{k}_{\shpa},x_{\perp}}$ with $i=1,2$ as the two orthonormal surface wavefunctions that satisfies \eqnref{eq:surf_modes_ori} at $\bsl{k}_{\shpa}$ with the boundary conditions. Orthonormality requires
\begin{equation}
\int_{-\infty}^0 d x_{\perp} v^{\dagger}_{i,\bsl{k}_{\shpa},x_{\perp}}v_{j,\bsl{k}_{\shpa},x_{\perp}}=\delta_{ij}\ .
\end{equation}
The creation operators of surface modes read
\begin{equation}
b^{\dagger}_{i,\bsl{k}_{\shpa}}=\int_{-\infty}^0 d x_{\perp} \Psi_{\bsl{k}_{\shpa},x_{\perp}}^{\dagger} v_{i,\bsl{k}_{\shpa},x_{\perp}}\ ,
\end{equation}
and the orthonormal condition of $v_{i,\bsl{k}_{\shpa},x_{\perp}}$ leads to the anti-commutation relations
\begin{equation}
\left\{b^{\dagger}_{i,\bsl{k}_{\shpa}}, b_{j,\bsl{k}'_{\shpa}}\right\}=\delta_{ij}\delta_{\bsl{k}_{\shpa} \bsl{k}'_{\shpa}}\ .
\end{equation}
The effective Hamiltonian for the surface zero modes can thus be expressed as
\begin{equation}
\label{eq:H_surf}
H_{surf}=E_{surf} \sum_{\bsl{k}_{\shpa}\in A}b^{\dagger}_{\bsl{k}_{\shpa}} b_{\bsl{k}_{\shpa}}\ ,
\end{equation}
where $A$ stands for the entire surface mode regions in the surface Brillouin zone, $E_{surf}=0$ and $b^{\dagger}_{\bsl{k}_{\shpa}}=(b^{\dagger}_{1,\bsl{k}_{\shpa}},b^{\dagger}_{2,\bsl{k}_{\shpa}})$.
Fermion parity operator will transform the $
b_{\bsl{k}_{\shpa}}$ operators as
$
b^{\dagger}_{\bsl{k}_{\shpa}}\rightarrow -b^{\dagger}_{\bsl{k}_{\shpa}}
$
and
$
b_{\bsl{k}_{\shpa}}\rightarrow -b_{\bsl{k}_{\shpa}}
$.
The 2D translation read
$
\hat{T}_{\bsl{x}_{\shpa}}b^{\dagger}_{\bsl{k}_{\shpa}}\hat{T}^{-1}_{\bsl{x}_{\shpa}}=e^{-i\bsl{k}_{\shpa}\cdot \bsl{x}_{\shpa}}b^{\dagger}_{\bsl{k}_{\shpa}}
$
and
$
\hat{T}_{\bsl{x}_{\shpa}}b_{\bsl{k}_{\shpa}}\hat{T}^{-1}_{\bsl{x}_{\shpa}}=e^{i\bsl{k}_{\shpa}\cdot \bsl{x}_{\shpa}}b_{\bsl{k}_{\shpa}}
$
.
Due to the TR symmetry, two orthonormal surface wavefunctions $v_{i,-\bsl{k}_{\shpa},x_{\perp}}$ at $-\bsl{k}_{\shpa}$
can be given by the linear combinations of $\mathcal{T}v^*_{i,\bsl{k}_{\shpa},x_{\perp}}$. Due to $\{ \mathcal{T} K, \chi \}=0$, $v_{i,-\bsl{k}_{\shpa},x_{\perp}}$ and $v_{i,\bsl{k}_{\shpa},x_{\perp}}$ have opposite chiral eigenvalues.
It means that $A_{\pm}$ can be related by $\bsl{k}_{\shpa}\rightarrow -\bsl{k}_{\shpa}$, where $A_{\pm}$ are the surface mode regions in the $\bsl{k}_{\shpa}$ space that are filled with the momenta of surface zero modes with chiral eigenvalue $\pm 1$, respectively.
Based on the same logic, $C_{3v}$ symmetries gives that $ v_{i, C_3 \bsl{k}_{\shpa},x_{\perp}}$ are linear combinations of $\widetilde{C}_3 v_{i, \bsl{k}_{\shpa},x_{\perp}}$ and $ v_{i, \Pi \bsl{k}_{\shpa},x_{\perp}}$ are linear combinations of $\widetilde{\Pi} v_{i, \bsl{k}_{\shpa},x_{\perp}}$.
Furthermore, since $\chi$ commutes with any operation in $C_{3v}$, $v_{i, C_3 \bsl{k}_{\shpa},x_{\perp}}$'s and $ v_{i, \Pi \bsl{k}_{\shpa},x_{\perp}}$'s have the same chiral eigenvalue as $v_{i, \bsl{k}_{\shpa},x_{\perp}}$, meaning that both $A_+$ and $A_-$ are $C_{3v}$ symmetric.
The representations of $\hat{\mathcal{T}}$, $\hat{C}_3$ and $\hat{\Pi}$ rely on the convention that we choose for $v_{i, \bsl{k}_{\shpa},x_{\perp}}$'s.
For convenience, we choose a special convention such that
\begin{equation}
\label{eq:v_conv}
\left\{
\begin{array}{l}
\mathcal{T}v_{i, \bsl{k}_{\shpa},x_{\perp}}^*=\sum_j v_{j, -\bsl{k}_{\shpa},x_{\perp}}(i\sigma_2)_{ji} \\
\widetilde{C}_3 v_{i, \bsl{k}_{\shpa},x_{\perp}}=
\sum_{j}v_{j, C_3\bsl{k}_{\shpa},x_{\perp}}(e^{-i \sigma_3 \frac{\pi}{3}})_{ji} \\
\widetilde{\Pi} v_{i, \bsl{k}_{\shpa},x_{\perp}}=
\sum_{j}v_{j, \Pi\bsl{k}_{\shpa},x_{\perp}}(-e^{-i \sigma_2 \frac{\pi}{2}})_{ji}
\end{array}
\right. \ .
\end{equation}
As a result, $b^{\dagger}_{\bsl{k}_{\shpa}}$ imitates a $j=1/2$ fermion:
\begin{equation}
\left\{
\begin{array}{l}
\hat{\mathcal{T}}b^{\dagger}_{\bsl{k}_{\shpa}}\hat{\mathcal{T}}^{-1}=
b^{\dagger}_{-\bsl{k}_{\shpa}}i\sigma_2 \\
\hat{C}_3 b^{\dagger}_{\bsl{k}_{\shpa}}\hat{C}_3^{-1}=
b^{\dagger}_{C_3\bsl{k}_{\shpa}}e^{-i \sigma_3 \frac{\pi}{3}} \\
\hat{\Pi} b^{\dagger}_{\bsl{k}_{\shpa}}\hat{\Pi}^{-1}=
b^{\dagger}_{\Pi \bsl{k}_{\shpa}}(-e^{-i \sigma_2 \frac{\pi}{2}})
\end{array}
\right. \ ,
\end{equation}
where  $\sigma_{1,2,3}$ are Pauli matrices for the double degeneracy of the surface modes.
And we can treat the double degeneracy of the surface modes as the pseudospin of the surface modes.
Since the PH symmetry is related with TR and chiral symmetries by $\chi=i \mathcal{T C^*}$, we have
\begin{equation}
\label{eq:PH_surf_v}
v_{i,-\bsl{k}_{\shpa},x_{\perp}}=\sum_{j=1}^2 \mathcal{C} v_{j,\bsl{k}_{\shpa},x_{\perp}}^*(\delta^{\chi}_{\bsl{k}_{\shpa}}\sigma_{2})_{ji}\ ,
\end{equation}
where $\delta^{\chi}_{\bsl{k}_{\shpa}}=\pm 1$ for $\bsl{k}_{\shpa}\in A_{\pm}$, $\chi v_{i,\bsl{k}_{\shpa},x_{\perp}}= \delta^{\chi}_{\bsl{k}_{\shpa}} v_{i,\bsl{k}_{\shpa},x_{\perp}}$, $\delta^{\chi}_{-\bsl{k}_{\shpa}}=-\delta^{\chi}_{\bsl{k}_{\shpa}}$ since $v_{i,\bsl{k}_{\shpa},x_{\perp}}$ and $v_{i,-\bsl{k}_{\shpa},x_{\perp}}$ have opposite chiral eigenvalues,
and $\delta^{\chi}_{R\bsl{k}_{\shpa}}=\delta^{\chi}_{\bsl{k}_{\shpa}}$ with $R\in C_{3v}$ since $v_{i,\bsl{k}_{\shpa},x_{\perp}}$ and $v_{i,R\bsl{k}_{\shpa},x_{\perp}}$ have the same chiral eigenvalue.
Furthermore, using $\Psi_{-\bsl{k}_{\shpa},x_{\perp}}^{\dagger}= \Psi_{\bsl{k}_{\shpa},x_{\perp}}^{T} \mathcal{C}$, we can get
\begin{equation}
b^{\dagger}_{-\bsl{k}_{\shpa}}=b^T_{\bsl{k}_{\shpa}}(\delta^{\chi}_{\bsl{k}_{\shpa}}\sigma_{2})\Leftrightarrow b^{\dagger}_{\bsl{k}_{\shpa}}(-\delta^{\chi}_{\bsl{k}_{\shpa}}\sigma_{2})=b^T_{-\bsl{k}_{\shpa}}\ .
\end{equation}
Thus, the PH symmetry gives rise to the following relation
\begin{eqnarray}
&&\left\{b_{i,\bsl{k}_{\shpa}}^{\dagger}, b_{j,\bsl{k}'_{\shpa}}^{\dagger}\right\}=
\left\{b_{i,\bsl{k}_{\shpa}}^{\dagger}, b_{i',-\bsl{k}'_{\shpa}} (\delta^{\chi}_{-\bsl{k}'_{\shpa}}\sigma_{2})_{i' j}\right\}\\
&&=
(\delta^{\chi}_{\bsl{k}_{\shpa}}\sigma_{2})_{ij}\delta_{\bsl{k}_{\shpa},- \bsl{k}'_{\shpa}}\ ,
\end{eqnarray}
which implies that only half the surface modes are actually physical due to the double counting of the BdG Hamiltonian.
In this case, we can treat the surfaces modes as two Majorana zero modes(MZMs) at each $\bsl{k}_{\shpa}$ as described below.
In general, the fermionic creation operator $b^{\dagger}_{i,\bsl{k}_{\shpa}}$ can be expressed as the linear combination of two Majorana operators:
$
b^{\dagger}_{i,\bsl{k}_{\shpa}}=\frac{1}{2}(\gamma_{i,\bsl{k}_{\shpa}}+i\widetilde{\gamma}_{i,\bsl{k}_{\shpa}})\ ,
$
where
\begin{equation}
\label{eq:exp_MFB}
\gamma_{i,\bsl{k}_{\shpa}}=b^{\dagger}_{i,\bsl{k}_{\shpa}}+b_{i,\bsl{k}_{\shpa}}\ ,
\end{equation}
and
$
\widetilde{\gamma}_{i,\bsl{k}_{\shpa}}=\frac{1}{i}(b^{\dagger}_{i,\bsl{k}_{\shpa}}-b_{i,\bsl{k}_{\shpa}})
$.
Due to \eqnref{eq:b_PH}, $\gamma_{i,\bsl{k}_{\shpa}}$ and $\widetilde{\gamma}_{i,\bsl{k}_{\shpa}}$ depend on each other by the relation
$
\gamma_{i,-\bsl{k}_{\shpa}}=-\delta^{\chi}_{\bsl{k}_{\shpa}}\sum_{j}\widetilde{\gamma}_{j,\bsl{k}_{\shpa}}(i\sigma_2)_{ji}\ .
$
Therefore, $\widetilde{\gamma}_{i,\bsl{k}_{\shpa}}$'s can be chosen to be redundant and we can treat the physical degrees of freedom as two MZMs at each $\bsl{k}_{\shpa}$, of which the Majorana operators are $\gamma_{i,\bsl{k}_{\shpa}}$.
And the $\gamma_{i,\bsl{k}_{\shpa}}$ operators satisfy the following anti-commutation relation:
\begin{eqnarray}
&&\{\gamma_{i,\bsl{k}_{\shpa}},\gamma_{j,\bsl{k}'_{\shpa}}\}=\nonumber\\
&&\{ b^{\dagger}_{i,\bsl{k}_{\shpa}},b^{\dagger}_{j,\bsl{k}_{\shpa}'}\}+\{ b^{\dagger}_{i,\bsl{k}_{\shpa}},b_{j,\bsl{k}_{\shpa}'}\}+\{ b_{i,\bsl{k}_{\shpa}},b^{\dagger}_{j,\bsl{k}_{\shpa}'}\}+\{ b_{i,\bsl{k}_{\shpa}},b_{j,\bsl{k}_{\shpa}'}\}\nonumber\\
&&=2\delta_{ij}\delta_{\bsl{k}_{\shpa},\bsl{k}'_{\shpa}}+(\delta^{\chi}_{\bsl{k}_{\shpa}}\sigma_{2})_{ij}\delta_{\bsl{k}_{\shpa},- \bsl{k}'_{\shpa}}+(\delta^{\chi}_{\bsl{k}_{\shpa}}\sigma_{2})_{ij}^*\delta_{\bsl{k}_{\shpa},- \bsl{k}'_{\shpa}}\nonumber\\
&& =2\delta_{ij}\delta_{\bsl{k}_{\shpa},\bsl{k}'_{\shpa}}\ .
\end{eqnarray}
Although the actual physical degrees of freedom are MZMs, we still use $b^{\dagger}_{\bsl{k}_{\shpa}}$ and $b_{\bsl{k}_{\shpa}}$ in the following for convenience.

\section{Projecting \eqnref{eq:Ht_mf} onto the surface to get \eqnref{eq:H_mf}}
\label{app:H_mf_c2b}
In this part, we will derive \eqnref{eq:H_mf} by projecting \eqnref{eq:Ht_mf} onto the surface.
First, we show the relation between the surface modes $b^{\dagger}$ and the Nambu bases $\Psi^{\dagger}$.
Due to the completeness of eigenstates of Hermitian operator, $\Psi^{\dagger}_{\bsl{k}_{\shpa},x_{\perp},\alpha,s}$ and $\Psi_{\bsl{k}_{\shpa},x_{\perp},\alpha,s}$ can be expressed in terms of eigenstates of \eqnref{eq:H_BdG_perp} for $x_{\perp}<0$ and $\bsl{k}_{\shpa}\in A$:
\begin{equation}
\left\{
\begin{array}{l}
\Psi^{\dagger}_{\bsl{k}_{\shpa},x_{\perp},\alpha,s}=\sum_{i} v^*_{i,\bsl{k}_{\shpa},x_{\perp},\alpha,s} b^{\dagger}_{i,\bsl{k}_{\shpa}}+\text{bulk modes}\\
\Psi_{\bsl{k}_{\shpa},x_{\perp},\alpha,s}=\sum_{i} v_{i,\bsl{k}_{\shpa},x_{\perp},\alpha,s} b_{i,\bsl{k}_{\shpa}}+\text{bulk modes}
\end{array}
\right. \ ,
\end{equation}
where $\alpha=e,h$ is the particle-hole index and $s=\pm\frac{3}{2},\pm\frac{1}{2}$.
Let us define $v_{\bsl{k}_{\shpa},x_{\perp}}$ as a $8\times 2$ matrix with $(\alpha,s)$ labeling the row and $i$ being the column index, and then the above relations can be expressed in the matrix version:
\begin{equation}
\label{eq:rel_b_psi}
\left\{
\begin{array}{l}
\Psi^{\dagger}_{\bsl{k}_{\shpa},x_{\perp}}= b^{\dagger}_{\bsl{k}_{\shpa}} v^{\dagger}_{\bsl{k}_{\shpa},x_{\perp}}+\text{bulk modes}\\
\Psi_{\bsl{k}_{\shpa},x_{\perp}}= v_{\bsl{k}_{\shpa},x_{\perp}} b_{\bsl{k}_{\shpa}}+\text{bulk modes}
\end{array}
\right. \ .
\end{equation}
In the matrix version, the symmetries of the surface eigenvectors become
\begin{equation}
\label{eq:v_conv_mat}
\left\{
\begin{array}{l}
\mathcal{T}v_{\bsl{k}_{\shpa},x_{\perp}}^*= v_{ -\bsl{k}_{\shpa},x_{\perp}} \mathcal{T}_b  \\
\widetilde{C}_3 v_{ \bsl{k}_{\shpa},x_{\perp}}=
v_{ C_3\bsl{k}_{\shpa},x_{\perp}}C_{3,b}\\
\widetilde{\Pi} v_{ \bsl{k}_{\shpa},x_{\perp}}=
v_{\Pi\bsl{k}_{\shpa},x_{\perp}}\Pi_b\\
v_{-\bsl{k}_{\shpa},x_{\perp}}= \mathcal{C} v_{\bsl{k}_{\shpa},x_{\perp}}^*\delta^{\chi}_{\bsl{k}_{\shpa}}\sigma_{2}\\
\chi v_{\bsl{k}_{\shpa},x_{\perp}}= \delta^{\chi}_{\bsl{k}_{\shpa}}v_{\bsl{k}_{\shpa},x_{\perp}}\\
\end{array}
\right. \ .
\end{equation}
If $\bsl{k}_{\shpa}$ is outside the surface mode regions, $\Psi^{\dagger}_{\bsl{k}_{\shpa},x_{\perp},\alpha,s}$ and $\Psi_{\bsl{k}_{\shpa},x_{\perp},\alpha,s}$ only contain bulk modes.

In the Nambu bases, \eqnref{eq:Ht_mf} reads
\begin{equation}
\label{eq:Ht_mf_Psi}
\widetilde{H}_{mf}=\frac{1}{2}\sum_{\bsl{k}_{\shpa}}^{A}\int_{-\infty}^{0} d x_{\perp}  \Psi^{\dagger}_{\bsl{k}_{\shpa},x_{\perp}}
\widetilde{h}(x_{\perp}) \Psi_{\bsl{k}_{\shpa},x_{\perp}}+const.\ ,
\end{equation}
where
\begin{equation}
\label{eq:ht}
\widetilde{h}(x_{\perp})
=
\left(
\begin{array}{cc}
\widetilde{M}(x_{\perp}) & \widetilde{D}(x_{\perp}) \\
\widetilde{D}^{\dagger}(x_{\perp}) & -\widetilde{M}^T(x_{\perp})
\end{array}
\right)\ .
\end{equation}
Using \eqnref{eq:rel_b_psi} and neglecting terms involving bulk modes, we can obtain \eqnref{eq:H_mf} with
$m(\bsl{k}_{\shpa})=\int_{-\infty}^0 dx_{\perp} v^{\dagger}_{\bsl{k}_{\shpa},x_{\perp}} \widetilde{h}(x_{\perp}) v_{\bsl{k}_{\shpa},x_{\perp}}$ being Hermitian. Due to the PH symmetry of $\widetilde{h}(x_{\perp})
$, i.e. $-\mathcal{C}\widetilde{h}^T(x_{\perp})\mathcal{C}^{\dagger}=\widetilde{h}(x_{\perp})
$, and $v_{\bsl{k}_{\shpa},x_{\perp}}$ in Eq. (\ref{eq:v_conv_mat}),
the obtained $m(\bsl{k}_{\shpa})$ is PH symmetric.
Only the TR odd part of $m(\bsl{k}_{\shpa})$, as well as $\widetilde{h}(x_{\perp})$, is allowed for the surface orders and thereby we only need to consider $\widetilde{h}(x_{\perp})$ satisfying $\mathcal{T}\widetilde{h}^*(x_{\perp})\mathcal{T}^{\dagger}=-\widetilde{h}(x_{\perp})$, which is equivalent to $\gamma \widetilde{M}^*(x_{\perp}) \gamma^{\dagger}=-\widetilde{M}(x_{\perp})$ and $\gamma \widetilde{D}^*(x_{\perp}) \gamma^{T}=-\widetilde{D}(x_{\perp})$.
Suppose $\widetilde{h}(x_{\perp})$ is the linear combination of $\widetilde{h}_i(x_{\perp})$ and $\widetilde{R} \widetilde{h}_i(x_{\perp})\widetilde{R}^{\dagger}=\sum_{j} f_{ij} \widetilde{h}_j(x_{\perp})$ with $f_{ij}\in \mathds{R}$, where the latter is equivalent to $R \widetilde{M}_i(x_{\perp}) R^{\dagger}=f_{ij} \widetilde{M}_j(x_{\perp})$ and $R \widetilde{D}_i(x_{\perp}) R^{T}=f_{ij} \widetilde{D}_j(x_{\perp})$, and  $R\in C_{3v}$.
According to the transformation of $v_{\bsl{k}_{\shpa},x_{\perp}}$ under $C_{3v}$ (\ref{eq:v_conv_mat}), we have $R_b \widetilde{m}_i (R^{-1}\bsl{k}_{\shpa}) R_b^{\dagger}=\sum_{j} f_{ij} \widetilde{m}_j (\bsl{k}_{\shpa})$, where $ \widetilde{m}_i (\bsl{k}_{\shpa})$ is the surface projection of $\widetilde{h}_i(x_{\perp})$.
Therefore, if $\widetilde{h}_i(x_{\perp})$, or equivalently $\widetilde{M}_i(x_{\perp})$ and $\widetilde{D}_i(x_{\perp})$, belongs to a certain IR of $C_{3v}$, the corresponding surface projection belongs to the same IR.

\section{Arcs of Majorana Zero Modes}
\label{app:MZM_arc}

\begin{figure}[t]
    \centering
    \includegraphics[width=\columnwidth]{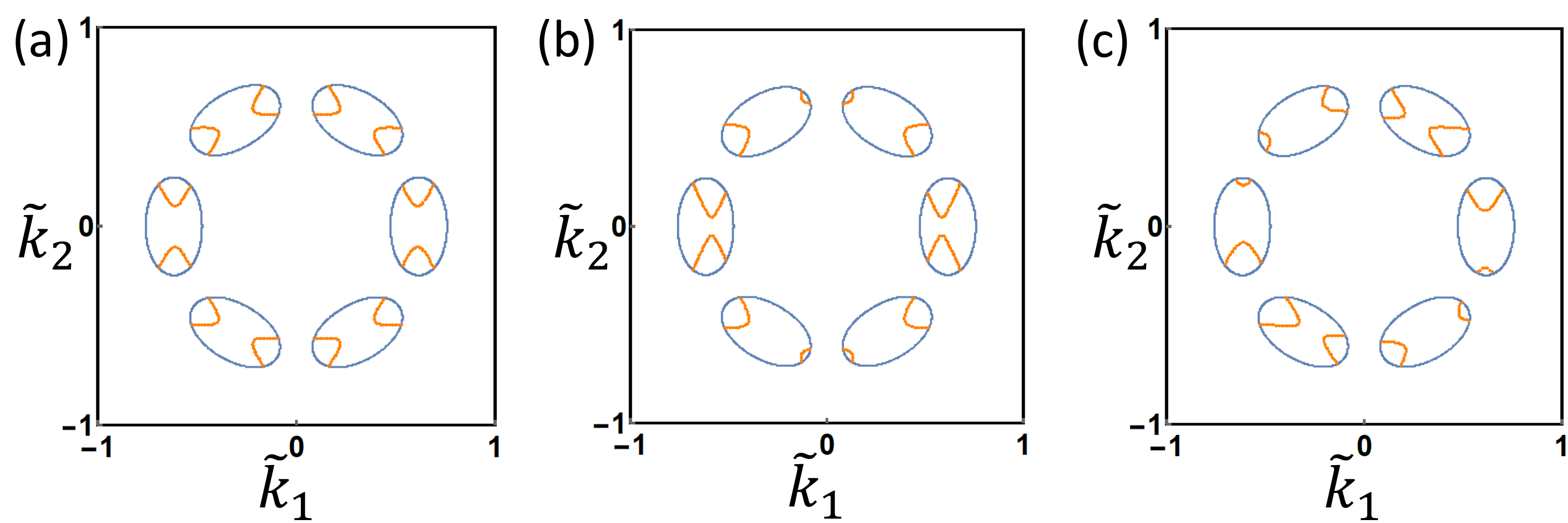}
    \caption{\label{fig:surf_MZM_arc} (a),(b) and (c) show the distribution of surface MZMs in the presence of $A_2$ surface translationally invariant order parameter without the $E$ order parameters, with the $\Pi$ anti-symmetric component of $E$ order parameters and with the $\Pi$ symmetric component of $E$ order parameters, respectively. Blue lines are the boundaries of surface mode regions shown in \figref{fig:surf_MFB} and one MZM exists on each point of orange lines.  $m_3/|\mu|=0.05$, $m_4/|\mu|=0.04$, $B_0\sqrt{2m/\mu}=0.8$, $B_1\sqrt{2m/\mu}=B_3\sqrt{2m/\mu}=1$ and $B_2\sqrt{2m/\mu}=-0.5$  are chosen for (a),(b) and (c), while $(m_{7,1}/|\mu|,m_{7,2}/|\mu|)=(0,0)$ for (a), $(m_{7,1}/|\mu|,m_{7,2}/|\mu|)=(0,0.05)$ for (b) and $(m_{7,1}/|\mu|,m_{7,2}/|\mu|)=(0.05,0)$ for (c). The non-zero values of $(m_{7,1}/|\mu|,m_{7,2}/|\mu|)$ indicate the existence of $E$ order. The values of all other parameters are the same as \figref{fig:surf_MFB}.}
\end{figure}

In this section, we will discuss the condition for the arcs of MZMs in the $\bsl{k}_\shpa$-space induced by order parameters. The analysis in \secref{sec:MF_order_MFB} only included orders that are uniform in each $A_{l_{\chi} l_c}$, and thereby the surface zero modes either exist or disappear at all $\bsl{k}_{\shpa}$ points in one $A_{l_{\chi} l_c}$ simultaneously.
If the momentum dependence of the orders within each $A_{l_{\chi} l_c}$ is considered, it is possible that MZMs exist at lines in the surface mode regions.
To illustrate that, we consider the $A_2$ order parameter to the linear order of momentum, which has no MZMs according to the analysis in \secref{sec:MF_order_MFB}.
To take into account the momentum dependence inside $A_{l_{\chi}, l_c}$, we define $\bsl{K}^{l_{\chi}, l_c}_{\shpa}$ to be the geometric center of $A_{l_{\chi}, l_c}$, and define $h_{A_2}^{l_{\chi}, l_c}(\bsl{q}_{\shpa})\equiv m_{A_2}(\bsl{q}_{\shpa}+\bsl{K}^{l_{\chi}, l_c}_{\shpa})$ with $\bsl{q}_{\shpa}\equiv \bsl{k}_{\shpa}-\bsl{K}^{l_{\chi}, l_c}_{\shpa}$.
Due to the odd mirror parity of $A_2$ order parameter and the $\Pi$ symmetry of $A_{+, 3}$,
$h_{A_2}^{+,3}(\bsl{q}_{\shpa})$ to the first order of $\bsl{q}_{\shpa}$ is
\begin{eqnarray}
\label{eq:h_RP3_A_2_k1}
&&h_{A_2}^{+,3}(\bsl{q}_{\shpa})=B_0 q_{\shpa,2}\sigma_0 + (-m_4 + B_1 q_{\shpa,1})\sigma_1+(- B_2 q_{\shpa,2})\sigma_2 \nonumber\\
&& + (m_3 +B_3 q_{\shpa,1})\sigma_3\ ,
\end{eqnarray}
where $K^{+,3}_{\shpa,2}=0$ is used.
In the following, we assume $B_{1,2,3,4} \neq 0$.
Using $C_{3v}$ and PH symmetries, we have $h_{A_2}^{+,1}(\bsl{q}_{\shpa})=C_{3,b} h_{A_2}^{+,3}(C_3^{-1}\bsl{q}_{\shpa})C_{3,b}^{\dagger}$, $h_{A_2}^{+,2}(\bsl{q}_{\shpa})=C_{3,b}^{\dagger} h_{A_2}^{+,3}(C_3\bsl{q}_{\shpa})C_{3,b}$, and $h_{A_2}^{-,l_c}(\bsl{q}_{\shpa})=-\sigma_2 [h_{A_2}^{+,l_c}(-\bsl{q}_{\shpa})]^T \sigma_2$.
As a result, the number of MZMs at $\bsl{k}_{\shpa}$ is the same as that at $C_3\bsl{k}_{\shpa}$, $\Pi \bsl{k}_{\shpa}$ and $-\bsl{k}_{\shpa}$, and thereby we only need to study the existence of MZMs in $A_{+, 3}$.
The eigenvalues of $h_{A_2}^{+,3}(\bsl{q}_{\shpa})$ are \begin{equation}
B_0 q_{\shpa,2}\pm \sqrt{(m_4 -B_1 q_{\shpa,1})^2 + (B_2 q_{\shpa,2})^2 + (m_3+ B_3 q_{\shpa,1})^2}\ .
\end{equation}
In the case where $-m_3/B_3=m_4/B_1$, two MZMs exist at $\bsl{q}_{\shpa}=(m_4/B_1, 0)$ if $(m_4/B_1, 0)\in A_{+, 3}$, and one MZM exists at every other point(in $A_{+, 3}$) on the straight line $(m_4/B_1, q_{\shpa,2})$  if $B_0^2- B_2 ^2=0$ or on the straight lines $(q_{\shpa,1}, \pm \sqrt{\frac{B_3^2+B_1^2}{B_0^2-B_2^2}} (m_4/B_1 - q_{\shpa,1}))$ if $B_0^2- B_2 ^2>0$.
In the case where $-m_3/B_3\neq m_4/B_1$, one MZM exists at every point on the part of the hyperbolas $(q_{\shpa,1}, \pm \sqrt{\frac{(m_4 - B_1 q_{\shpa,1})^2+(m_3 + B_3 q_{\shpa,1})^2}{B_0^2-B_2^2})}$ that is in $A_{+, 3}$ if $B_0^2- B_2 ^2>0$.
If none of the conditions listed above are satisfied, no MZMs exist.
As an example, \figref{fig:surf_MZM_arc}a shows the surface Majorana arcs for $B_0^2- B_2 ^2>0$ and $-m_3/B_3\neq m_4/B_1$, where only one MZM exists at each point of the arcs and the distribution of MZMs has $C_{3v}$ and PH symmetries as mentioned before.
In the plot, we assume only surface order is formed and the bulk nodal lines as well as the boundaries of surface mode regions do not change.
Such distribution of Majorana arcs is possible to be generated by surface FM along the $(111)$ direction since it is an $A_2$ order parameter.

Next we consider how the $E$ order parameter changes the distribution of Majorana arcs.
Suppose the surface Majorana arcs exist for the $A_2$ order which is given by surface FM in the $(111)$ direction.
In this case, the presence of the small $E$ order parameter can be achieved by tuning the surface magnetic moment slightly away from the $(111)$ direction with a weak external magnetic field, which can change the distribution of the surface Majorana arcs.
To illustrate that, we add only the momentum independent $E$ order parameter $\bsl{m}_7\cdot\bsl{N}_7$ to the $A_2$ order $h_{A_2}^{l_\chi,l_c}(\bsl{q}_{\shpa})$ for simplicity.
If the magnetic moment is tilted to $(11\bar{2})$ direction, then the system still has odd $\Pi$ parity, meaning that $m_{7,1}=0$.
In this case, the $C_3$ symmetry of the distribution of surface Majorana arc is broken while its $\Pi$ symmetry is preserved, which is exactly shown in \figref{fig:surf_MZM_arc}b.
If the magnetic moment is tilted to $(\bar{1}10)$ direction, then the extra term should be $\Pi$ symmetric, meaning that $m_{7,2}=0$.
As a result, the entire $C_{3v}$ symmetry of the surface Majorana arc distribution is broken, which matches \figref{fig:surf_MZM_arc}c.

\section{More Details on Impurity Effect}
\label{app:H_d_bases}
In this section, we will provide more details on the impurity effect of SMFBs.

\subsubsection{Order Parameters in $\mathbf{r}_{\shpa}$ space}
In this part, we will discuss the transformation of order parameters from the $\bsl{k}_{\shpa}$ space to the $\bsl{r}_{\shpa}$ space.
Let us consider the general order parameters that are independent of $\bsl{k}_{\shpa}$ in each $A_{l_\chi, l_c}$, i.e. \eqnref{eq:H_mf} with $m(\bsl{k}_{\shpa})$ having the form \eqnref{eq:m_unif_gen}.
Using \eqnref{eq:b_r_k} and \eqnref{eq:appro_local_d}, we have
\begin{equation}
H_{mf}=\frac{1}{2}\int dr^2_{\shpa} d^{\dagger}_{\bsl{r}_{\shpa}} M d_{\bsl{r}_{\shpa}}\ ,
\end{equation}
with $M_{l_\chi l_\chi',l_c l_c',i i'}=\sum_{l=0}^3 f^{l_\chi,l_c}_{l}(\sigma_l)_{ii'}\delta_{l_\chi l_\chi'} \delta_{l_c l_c'}$. $f^{l_\chi,l_c}_{l}$'s for different $l_\chi,l_c$ are given by $1$ or $\delta^{\alpha}_{\bsl{k}_{\shpa}}$ with $\alpha=\chi, (E_1,\pm), (E_2,\pm)$.
Specifically, we have
\begin{eqnarray}
&& 1 =\sum_{l_\chi, l_c} (\rho_0)_{l_\chi l_\chi} (\Lambda_{1})_{l_c l_c}\delta^{l_\chi,l_c}_{\bsl{k}_{\shpa}}\nonumber\\
&& \delta^{\chi}_{\bsl{k}_{\shpa}} =\sum_{l_\chi, l_c} (\rho_3)_{l_\chi l_\chi} (\Lambda_{1})_{l_c l_c}\delta^{l_\chi,l_c}_{\bsl{k}_{\shpa}}\nonumber\\
&& \delta^{E_1,+}_{\bsl{k}_{\shpa}} =\sum_{l_\chi, l_c} (\rho_0)_{l_\chi l_\chi} (\Lambda_{4,1})_{l_c l_c}\delta^{l_\chi,l_c}_{\bsl{k}_{\shpa}}\nonumber\\
&& \delta^{E_1,-}_{\bsl{k}_{\shpa}} =\sum_{l_\chi, l_c} (\rho_3)_{l_\chi l_\chi} (\Lambda_{4,1})_{l_c l_c}\delta^{l_\chi,l_c}_{\bsl{k}_{\shpa}}\nonumber\\
&& \delta^{E_2,+}_{\bsl{k}_{\shpa}} =\sum_{l_\chi, l_c} (\rho_0)_{l_\chi l_\chi} (\Lambda_{4,2})_{l_c l_c}\delta^{l_\chi,l_c}_{\bsl{k}_{\shpa}}\nonumber\\
&& \delta^{E_2,-}_{\bsl{k}_{\shpa}} =\sum_{l_\chi, l_c} (\rho_3)_{l_\chi l_\chi} (\Lambda_{4,2})_{l_c l_c}\delta^{l_\chi,l_c}_{\bsl{k}_{\shpa}}\ ,
\end{eqnarray}
where that all matrices involved are diagonal due to translation symmetry.
Using the above correspondence, \tabref{tab:N} and \eqnref{eq:H_mf_A1}-\ref{eq:H_mf_E}, we can get
\begin{equation}
H_{mf}^{\alpha}=\frac{1}{2}\int d^2 \bsl{r}_{\shpa} d^{\dagger}_{\bsl{r}_{\shpa}} M_\alpha d_{\bsl{r}_{\shpa}}+const.\ ,
\end{equation}
where $\alpha=A_1,A_2,E$,
\begin{equation}
\label{eq:M_A1}
M_{A_1}=m_1\rho_3\otimes \Lambda_1 \otimes \sigma_0+m_2(-\rho_0\otimes \Lambda_{4,1} \otimes \sigma_2+\rho_0\otimes \Lambda_{4,2} \otimes \sigma_1)\ ,
\end{equation}
\begin{equation}
\label{eq:M_A2}
M_{A_2}=m_3 \rho_0\otimes \Lambda_1 \otimes \sigma_3+m_4 (\rho_0\otimes\Lambda_{4,2}\otimes\sigma_2+\rho_0\otimes\Lambda_{4,1}\otimes\sigma_1)\ ,
\end{equation}
and
\begin{eqnarray}
\label{eq:M_E}
&&M_E=m_{5,1} \rho_3\otimes\Lambda_{4,1}\otimes\sigma_0+m_{5,2}\rho_3\otimes\Lambda_{4,2}\otimes\sigma_0\nonumber\\
&&+m_{6,1}(-\rho_0\otimes\Lambda_{4,2}\otimes\sigma_3)+m_{6,2} (\rho_0\otimes\Lambda_{4,1}\otimes \sigma_3)\nonumber\\
&&+m_{7,1}(-\rho_0\otimes\Lambda_1\otimes\sigma_2)+m_{7,2}(\rho_0\otimes\Lambda_1\otimes\sigma_1)\nonumber\\
&&+m_{8,1}(\rho_0\otimes\Lambda_{4,1}\otimes\sigma_2+\rho_0\otimes\Lambda_{4,2}\otimes\sigma_1)\nonumber\\
&&+m_{8,2}(\rho_0\otimes\Lambda_{4,1}\otimes\sigma_1-\rho_0\otimes\Lambda_{4,2}\otimes\sigma_2)\ .
\end{eqnarray}
According to \tabref{tab:IR_C3v_TR_PH_chi}, \eqnref{eq:M_A1}, \eqnref{eq:M_A2} and \eqnref{eq:M_E} are the most general PH symmetric uniform order parameters for the $A_1$, $A_2$ and $E$ IRs.


\subsubsection{Verification of LDOS Peaks for Translational Invariant Order Parameters with $d$ Bases}
The purpose for this section is to re-derive the distribution of LDOS peaks from the symmetry aspect of the order parameters in \eqnref{eq:M_A1}-\ref{eq:M_E} with the $d$ bases and establish the formalism that can be generalized to the case with charge/magnetic impurities.
Since the position $\bsl{r}_{\shpa}$ is now approximately a good quantum number, the number of LDOS peaks is directly determined by the number of different eigenvalues of $M_{\alpha}$.
It means that the numbers of LDOS peaks far away from impurities should be typically 1,4,2 and 12 for no order parameters, the $A_1$ order parameter, the $A_2$ order parameter and the $E$ order parameter, respectively, as indicated in \secref{sec:MF_order_MFB}.
12 LDOS peaks for the $E$ order parameter are justified by the fact that $M_{\alpha}$'s are all $12\times 12 $ matrices with 12 eigenvalues and the $E$ order parameter typically has no symmetries to ensure any degeneracy.
To discuss $A_1$ and $A_2$ order parameters, we again transform all the symmetry operators to the eigenbases of $C_{3,d}$ as discussed in the main text.
By choosing the same convention (\ref{eq:C3t_d},\ref{eq:Pit_d}) in the main text, the representations of the symmetry operations other than $\hat{C}_{3}$ and $\hat{\Pi}$ are
\begin{equation}
\widetilde{U}_{T}
=
\left(
\begin{array}{ccc}
 & \mathds{1}_4 & \\
& & \mathds{1}_4\\
\mathds{1}_4 & & \\
\end{array}
\right)\ ,
\end{equation}
\begin{equation}
\widetilde{\mathcal{C}}_{d}
=
\left(
\begin{array}{ccc}
 &  & U_c\\
& U_c & \\
U_c & & \\
\end{array}
\right)
\end{equation}
with
\begin{equation}
U_c=
\left(
\begin{array}{cccc}
 0 & 0 & 0 & i \\
 0 & 0 & -i & 0 \\
 0 & -i & 0 & 0 \\
 i & 0 & 0 & 0 \\
\end{array}
\right)\ ,
\end{equation}
and
\begin{equation}
\widetilde{\chi}_{d}
=
\left(
\begin{array}{ccc}
U_\chi &  & \\
 & U_\chi & \\
 & & U_\chi\\
\end{array}
\right)
\end{equation}
with
\begin{equation}
U_\chi=
\left(
\begin{array}{cccc}
 -1 & 0 & 0 & 0 \\
 0 & -1 & 0 & 0 \\
 0 & 0 & 1 & 0 \\
 0 & 0 & 0 & 1 \\
\end{array}
\right)\ ,
\end{equation}
where $\widetilde{R}$ means the matrix form of $R$ in the $C_{3,d}$ eigenbases and $U_T$ is defined such that $M$ is diagonal for $l_c$ index if and only if $[M,U_T]=0$.
The $A_1$ order parameter satisfies $[M_{A_1},C_{3,d}]=[M_{A_1},U_T]=0$.
Due to the commutation relation with $C_{3,d}$, $\widetilde{M}_{A_1}$ should be block-diagonal and written as $\widetilde{M}_{A_1}=\text{diag}(h_1,h_2,h_3)$, where $h_{1,2,3}$ are Hermitian $4\times 4$ matrices.
Furthermore, due to the commutation relation with $U_T$, we requires $h_1=h_2=h_3$, which leads to the three-fold degeneracy of each eigenvalues.
As a result, $M_{A_1}$ has typically 4 LDOS peaks.
The $A_2$ order parameter satisfies not only $[M_{A_2},C_{3,d}]=[M_{A_2},U_T]=0$ but also $[M_{A_2}, \Pi_d \mathcal{C}_d K]=0$, in which we have $(\Pi_d \mathcal{C}_d K)^2=-1$.
The former leads to $\widetilde{M}_{A_2}=\text{diag}(h_1,h_1,h_1)$ as mentioned above, while  $\Pi_d \mathcal{C}_d M^* \mathcal{C}_d^{\dagger} \Pi_d^{\dagger}=M$ results in $U_{\Pi} U_c h_1^* U_c^{\dagger} U_\Pi^{\dagger}=h_1$.
Thereby, each eigenvalues of $h_1$ have double degeneracy due to $U_{\Pi} U_c (U_{\Pi} U_c)^*=-1$.
As a result, all eigenvalues of $M_{A_2}$ have six-fold degeneracy and the $A_2$ order parameter typically has 2 peaks.
In addition, $M_{\alpha}$'s are PH symmetric, which guarantees that LDOS peaks are symmetric with respect to zero energy.

\subsubsection{Derivation of \eqnref{eq:H_imp_d} and the Symmetry Properties}
In this part, we will derive \eqnref{eq:H_imp_d} and discuss the corresponding symmetry properties.
The surface impurity Hamiltonian that we consider has the general form
\begin{equation}
\label{eq:H_imp_c}
H_V=\int d^3 r c^{\dagger}_{\bsl{r}} V(\bsl{r}) c_{\bsl{r}}\ ,
\end{equation}
where the position of the impurity is at $\bsl{r}=0$ (certainly on the $x_{\perp}=0$ surface) and $V(\bsl{r})^{\dagger}=V(\bsl{r})$ decays fast away from $\bsl{r}=0$.
First we express \eqnref{eq:H_imp_c} in the Nambu bases as
\begin{eqnarray}
&&H_V=\frac{1}{2}\int d^2 r_{\shpa} \int d x_{\perp} \frac{1}{S_{\shpa}}\sum_{\bsl{k}_{\shpa},\bsl{k}_{\shpa}'}e^{-i \bsl{k}_{\shpa}\cdot \bsl{r}_{\shpa}+i \bsl{k}_{\shpa}'\cdot \bsl{r}_{\shpa}}\nonumber\\
&&\Psi^{\dagger}_{\bsl{k}_{\shpa},x_{\perp}}
\widetilde{V}(\bsl{r})
 \Psi_{\bsl{k}_{\shpa}',x_{\perp}}
 + const.\ ,
\end{eqnarray}
where
\begin{equation}
\widetilde{V}(\bsl{r})
=\left(
\begin{array}{cc}
V(\bsl{r}) & \\
 & -V^*(\bsl{r})\\
\end{array}
\right)\ ,
\end{equation}
and
$
\Psi_{\bsl{r}}^{\dagger}=\frac{1}{\sqrt{S_{\shpa}}}\sum_{\bsl{k}_{\shpa}}e^{-i \bsl{k}_{\shpa}\cdot \bsl{r}_{\shpa}}\Psi_{\bsl{k}_{\shpa},x_{\perp}}^{\dagger}
$
is used. Using \eqnref{eq:rel_b_psi}, we only keep terms that involve surface modes and assume $v_{\bsl{k}_{\shpa},x_{\perp}}\approx v_{\bsl{K}_{\shpa}^{l_\chi, l_c},x_{\perp}}$ for all $\bsl{k}_{\shpa}\in A_{l_\chi, l_c}$ and all ${l_\chi, l_c}$. This leads to \eqnref{eq:H_imp_d}
with
\begin{equation}
[M_V(\bsl{r}_{\shpa})]_{l_\chi l_\chi',l_c l_c', i i'}=\int_{-\infty}^0 d x_{\perp} v^{\dagger}_{i,\bsl{K}^{l_\chi,l_c}_{\shpa},x_{\perp}} \widetilde{V}(\bsl{r}) v_{i',\bsl{K}^{l_\chi',l_c'}_{\shpa},x_{\perp}}\ .
\end{equation}
Since $V^{\dagger}(\bsl{r})=V(\bsl{r})$, we have $M_V^{\dagger}(\bsl{r}_{\shpa})=M_V(\bsl{r}_{\shpa})$.
Due to
\begin{equation}
\sum_{l_\chi',l_c',i'}[\mathcal{C}_d]_{l_{\chi} l_\chi',l_c l_c',i i'} v_{i',\bsl{K}_{\shpa}^{l_\chi', l_c'},x_{\perp}}=\mathcal{C}  v^*_{i,\bsl{K}_{\shpa}^{l_\chi, l_c},x_{\perp}}\ ,
\end{equation}
$M_V(\bsl{r}_{\shpa})$ is PH symmetric, written as
\begin{equation}
-\mathcal{C}_d M_V^T(\bsl{r}_{\shpa}) \mathcal{C}_d^{\dagger}=M_V(\bsl{r}_{\shpa})\ .
\end{equation}
Due to
\begin{equation}
\sum_{l_\chi',l_c',i'}[\mathcal{T}_d]_{l_{\chi} l_\chi',l_c l_c',i i'} v_{i',\bsl{K}_{\shpa}^{l_\chi', l_c'},x_{\perp}}=\mathcal{T}^T  v^*_{i,\bsl{K}_{\shpa}^{l_\chi, l_c},x_{\perp}}\ ,
\end{equation}
$M_V(\bsl{r}_{\shpa})$ has the same TR properties as $\widetilde{V}(\bsl{r})$:
\begin{eqnarray}
&&[\mathcal{T}_d M_V^*(\bsl{r}_{\shpa}) \mathcal{T}_d^{\dagger}]_{l_{\chi} l_\chi',l_c l_c',i i'}=\nonumber\\
&&\int_{-\infty}^0 d x_{\perp} v^{\dagger}_{i,\bsl{K}^{l_\chi,l_c}_{\shpa},x_{\perp}} \mathcal{T}\widetilde{V}^*(\bsl{r}) \mathcal{T}^{\dagger}v_{i',\bsl{K}^{l_\chi',l_c'}_{\shpa},x_{\perp}}\ .
\end{eqnarray}
Similarly, due to
\begin{equation}
\sum_{l_\chi',l_c',i'}[R_d]_{l_{\chi} l_\chi',l_c l_c',i i'} v^{\dagger}_{i',\bsl{K}_{\shpa}^{l_\chi', l_c'},x_{\perp}}= v^{\dagger}_{i,\bsl{K}_{\shpa}^{l_\chi, l_c},x_{\perp}}\widetilde{R}\ ,
\end{equation}
$M_V(\bsl{r}_{\shpa})$ has the same $C_{3v}$ properties as $\widetilde{V}(\bsl{r})$:
\begin{eqnarray}
&&[\mathcal{R}_d M_V(\bsl{r}_{\shpa}) \mathcal{R}_d^{\dagger}]_{l_{\chi} l_\chi',l_c l_c',i i'}=\nonumber\\
&&\int_{-\infty}^0 d x_{\perp} v^{\dagger}_{i,\bsl{K}^{l_\chi,l_c}_{\shpa},x_{\perp}} \widetilde{R}\widetilde{V}(\bsl{r})\widetilde{R}^{\dagger}v_{i',\bsl{K}^{l_\chi',l_c'}_{\shpa},x_{\perp}}\ ,
\end{eqnarray}
where $R\in C_{3v}$.
Furthermore, since $\widetilde{V}(\bsl{r})$ behaves the same as $V(\bsl{r})$, the TR and $C_{3v}$ properties of $M_{V}(\bsl{r}_{\shpa})$ are the same as those of $V(\bsl{r})$.

For a charge impurity, $V(\bsl{r})=V_c(\bsl{r})\mathds{1}_{4\times 4}$ with $V_c(\bsl{r})$ a real scalar function.
In this case, $V_c(\bsl{r})\mathds{1}_{4\times 4}$ has TR symmetry $\gamma (V_c(\bsl{r})\mathds{1}_{4\times 4})^* \gamma^{\dagger}=V_c(\bsl{r})\mathds{1}_{4\times 4}$ and satisfies $R (V_c(\bsl{r})\mathds{1}_{4\times 4}) R^{\dagger}=V_c(\bsl{r})\mathds{1}_{4\times 4}$ with $R\in C_{3v}$. As a result, Hermitian and PH symmetric $M_V(\bsl{r})$ has TR symmetry $\mathcal{T}_d M^*_{V}(\bsl{r}_{\shpa})\mathcal{T}_d^{\dagger}=M_V(\bsl{r}_{\shpa})$ and satisfies $R_d M_V(\bsl{r}_{\shpa}) R_d^{\dagger}=M_V(\bsl{r}_{\shpa})$ with $R\in C_{3v}$.
Combining TR and PH symmetries, we have chiral symmetry for $M_V(\bsl{r}_{\shpa})$, i.e. $\chi_d M_V(\bsl{r}_{\shpa}) \chi_d^{\dagger}=-M_V(\bsl{r}_{\shpa})$.
By defining $M_c=M_V(\bsl{r}_\shpa =0)$, the symmetry properties of $M_c$ can be directly obtained.

For a magnetic impurity, we choose the magnetic moment of the impurity to be perpendicular to the surface and couple to the electron spin locally, i.e. choosing $V(\bsl{r})=V_m(\bsl{r}) \bsl{e}_{\perp}\cdot \bsl{J}$ with $V_m(\bsl{r})$ a real scalar function and $\bsl{e}_{\perp}= (1,1,1)/\sqrt{3}$. In this case, $V_m(\bsl{r}) \bsl{e}_{\perp}\cdot \bsl{J}$ is TR odd $\gamma (V_m(\bsl{r}) \bsl{e}_{\perp}\cdot \bsl{J})^* \gamma^{\dagger}=-V_m(\bsl{r}) \bsl{e}_{\perp}\cdot \bsl{J}$, and satisfies $C_3 (V_m(\bsl{r}) \bsl{e}_{\perp}\cdot \bsl{J}) C_3^{\dagger}=V_m(\bsl{r}) \bsl{e}_{\perp}\cdot \bsl{J}$ and  $\Pi (V_m(\bsl{r}) \bsl{e}_{\perp}\cdot \bsl{J}) \Pi^{\dagger}=-V_m(\bsl{r}) \bsl{e}_{\perp}\cdot \bsl{J}$.
As a result, the Hermitian and PH symmetric $M_V(\bsl{r}_{\shpa})$ has TR antisymmetry $\mathcal{T}_d M_V^*(\bsl{r}_{\shpa})\mathcal{T}_d^{\dagger}=-M_V(\bsl{r}_{\shpa})$, and satisfies $C_{3,d} M_V(\bsl{r}_{\shpa})C_{3,d}^{\dagger}=M_V(\bsl{r}_{\shpa})$ and $\Pi_{d} M_V(\bsl{r}_{\shpa})\Pi_{d}^{\dagger}=-M_V(\bsl{r}_{\shpa})$.
By defining $M_m=M_V(\bsl{r}_\shpa=0)$, the symmetry properties of $M_m$ can be obtained.

In \figref{fig:LDOS},  $V_c(\bsl{r})/|\mu|=2/(|\bsl{r}|\sqrt{2m\mu}+0.02)^2$ if the charge impurity is considered, and $V_m(\bsl{r})/|\mu|=5 e^{x_{\perp} \sqrt{2m\mu}/2}\theta(|\bsl{r}_{\shpa,0}|-|\bsl{r}_{\shpa}|)$ with $|\bsl{r}_{\shpa}|<|\bsl{r}_{\shpa,0}|$ if the magnetic impurity is considered.

\end{appendices}

%

\end{document}